\documentclass{ws-ijmpa}
\usepackage{tikz}
\usetikzlibrary{decorations.pathreplacing}
\usepackage{braket}
\usepackage{mathtools}
\usepackage{enumitem, bbm}
\usepackage{lipsum}
\usepackage{comment}
\usepackage[super]{cite}
\usepackage{xcolor,enumitem}
\usepackage[verbose,hypertexnames=false, hidelinks]{hyperref}
\hypersetup{colorlinks=false,allbordercolors=blue,pdfborderstyle={/S/U/W 1}}
\usepackage{tikz}
\usetikzlibrary{positioning, arrows.meta, calc}

\begin{document}

\markboth{Renata Ferrero}{Asymptotic Safety and Canonical Quantum Gravity}

%
\catchline{}{}{}{}{}
%

\title{Asymptotic Safety and Canonical Quantum Gravity}

\author{Renata Ferrero
}

\address{Institute for Quantum Gravity, Friedrich-Alexander-Universität Erlangen-Nürnberg, \\
	Staudtstr. 7, 91058 Erlangen, Germany\\
renata.ferrero@fau.de}

\maketitle


\begin{abstract}
In the context of gravity the   Lagrangian  and Hamiltonian formalisms have been developed largely independently, emphasizing   renormalization and quantization, respectively. The formalisms use a different methodology to distinguish between gauge and physical degrees of freedom. In this review we analyze the connection between the Asymptotically Safe and Canonical Quantum Gravity approaches. Based on the Hamiltonian formulation, the Canonical Quantum Gravity approach inherently provides a natural framework for defining observables.  This serves as the foundation for constructing the generating functional of the $n$-point correlation functions of physical degrees of freedom. By means of background-independent, non-perturbative renormalization methods well-established in the  Lagrangian framework and typically employed in Asymptotic Safety, the resulting generating functional can be handled. In particular, we employ the Functional Renormalization Group  to regularize the path integral and to compute the  flow connecting the bare theory in the ultraviolet  with the effective infrared theory.  An important advantage of this approach is that it establishes an explicit, systematic relation between the quantization procedure and the systematics of quantum field theory-based renormalization group methods. More importantly, this synthesis not only bridges canonical and covariant approaches but also paves the way for a consistent and predictive quantum theory of gravity grounded in physically meaningful, gauge-invariant observables.
\end{abstract}



\section{Introduction}	
A theory of quantum gravity is essential to achieve a deeper and more unified understanding of the universe across all  scales. Currently, physics relies on two foundational yet fundamentally different theories: General Relativity (GR) and Quantum Field Theory (QFT). GR describes gravity and the structure of spacetime at large distances, while QFT allows precision computations describing the behavior of matter and forces at the subatomic level.

Both theories are remarkably successful within their respective domains and have been thoroughly validated by  observations \cite{LIGOScientific:2021sio,ParticleDataGroup:2024cfk}. However, they are based on distinct conceptual foundations and mathematical frameworks, and  encounter issues when applied simultaneously at all scales. Particularly important is their interplay in extreme regimes such as in the vicinity of singularities in black holes or in the early universe, where both quantum and gravitational effects are assumed to play a role.

This theoretical divide suggests that our current understanding of nature is incomplete. Since the other fundamental forces are described using quantum principles, it is natural to expect that gravity should also be incorporated into a quantum framework \cite{Kiefer:2007ria}. 
There are multiple approaches to developing  a quantum theory of gravity, each reflecting different strategies and conceptual priorities \cite{Armas:2021yut,Bambi:2023jiz,Buoninfante:2024yth}. One of the earliest and most influential attempts emerged from the canonical formalism. Here gravity is quantized by applying techniques analogous to those used in quantum mechanics \cite{Dirac:1949cp,Dirac:1950pj,Dirac:1958sc,Anderson:1951ta,Bergmann:1954tc,Komar:1958wp,Bergmann:1960wb}. This led to the formulation of the Wheeler--DeWitt equation \cite{Wheeler:1957mu,wheeler1962geometrodynamics,DeWitt:1967yk,DeWitt:1967ub,DeWitt:1967uc}, acting as the  Schrödinger equation of the universe. It describes the quantum state of the entire gravitational field, but without an explicit time parameter, highlighting the so-called ``problem of time" in quantum gravity.
Building on this foundation, significant progress has been made in addressing key conceptual and technical challenges, particularly in understanding the nature of time and constructing a well-defined Hamiltonian formulation of quantum gravity \cite{Thiemann:2007pyv, Thiemann:2023zjd}. A prominent development in this direction is Loop Quantum Gravity (LQG) \cite{Ashtekar:2004eh,Rovelli:2008zza,Ashtekar:2021kfp,Rovelli:2014ssa}, which applies a background-independent quantization of the gravitational field in terms of Ashtekar--Barbero variables.

In parallel to canonical quantization approaches, the development of QFT techniques inspired alternative methods to incorporate gravity into a quantum framework. One outstanding example is the Asymptotic Safety (AS) program \cite{Percacci:2017fkn,Reuter:2019byg,Niedermaier:2006wt,Reuter:2012id}. It seeks to construct a consistent and predictive quantum theory of gravity using concepts rooted in the theory of critical phenomena.
At the heart of this approach is the realization of a non-trivial ultraviolet (UV) fixed point in the renormalization group (RG) flow of gravitational couplings \cite{Weinberg:1976xy,Weinberg:1980gg}. Unlike perturbative methods, which fail to achieve  UV completeness and  face challenges with the renormalizability of the Einstein--Hilbert action, Asymptotic Safety employs non-perturbative RG techniques, particularly the Functional Renormalization Group  (FRG) \cite{Dupuis:2020fhh,Saueressig:2023irs}.
A key feature of this framework is that it complies with the principle of background independence, meaning that the procedure does not rely on a fixed spacetime background. Instead, it treats the spacetime metric dynamically, following principles of GR \cite{Kiefer:2005uk,Giulini:2006yg,Giulini:2006xi,Kiefer:2013jqa}.

Historically, these two approaches to quantum gravity have been developed largely independently, with each being grounded in distinct conceptual foundations and mathematical tools. 
However, a common challenge across these  approaches has been the definition and identification of observables. In gravity, the diffeomorphism invariance of the theory complicates the definition of local observables, as the theory lacks a fixed background structure \cite{Bergmann:1949zz,Bergmann:1954tc,Bergmann:1972ud,Bergmann:1960wb}. This issue persists in quantum gravity, where the absence of a fixed background spacetime  makes it difficult to define quantities that can be measured or observed in a traditional sense. Consequently, the need to establish observables has become increasingly important, prompting advancements from different quantum gravity approaches to seek common ground and develop methods that can bridge their conceptual differences. 
This convergence is reflected in efforts to define and utilize relational observables \cite{Rovelli:1989jn,Rovelli:1990pi,Rovelli:1990ph,Rovelli:2001bz,Dittrich:2005kc,Dittrich:2004cb,Tambornino:2011vg}, quantities defined through the relations between physical entities rather than with respect to a fixed background spacetime.

In this review, we highlight  the commonalities and differences between the two approaches. Conceptual differences  stem from the differences between covariant and Hamiltonian formulations and the related challenges to embed a quantized theory of gravity within a QFT framework. One particular point of separation is the construction of observable quantities: in the Hamiltonian setup the quantization procedure can be performed after having identified a physical time direction, while in the covariant QFT setting the  path integral is equipped with a gauge fixing. We will discuss under which assumptions these two procedures can be related explicitly.

Gravity presents a unique challenge, particularly due to the role of time in GR. After having identified physical degrees of freedom within the relational framework, we will show how the Hamiltonian formalism provides a natural quantum mechanical definition of observables suited for gravity, though it lacks a thorough method to access the effective, infrared (IR) regime.
In contrast, within the Asymptotic Safety program   a fully systematic notion of diffeomorphism-invariant observables has yet to be established. While running couplings and fixed points can be computed, physical quantities like scattering amplitudes remain difficult to define within a fully background-independent setting.
 Once diffeomorphism-invariant  observables are established, these   could  be analyzed using the RG  machinery and linked to the universal properties of the theory in a predictive way.

\bigskip
This review is structured as follows.  In Section \ref{sec:2} we discuss  the  principles shared by Asymptotic Safety  and Canonical Quantum Gravity (CQG). In Section \ref{sec:3} we discuss the motivations in connecting these two approaches, with the goal of defining candidate observables for quantum gravity. We further  review  the conceptual differences between the Hamiltonian and the Lagrangian formalism, as used in  CQG and AS. Sections \ref{sec:4} and \ref{sec:5}  focus on a detailed introduction to AS and the FRG, and CQG and its path integral formulation. Section \ref{sec:4} or \ref{sec:5} may be skipped by readers already familiar with AS  and CQG, respectively. In Section \ref{sec:6} we exemplify, using two classes of models, how the AS and CQG approaches can concretely be merged, highlighting the novelties of the resulting   asymptotically safe canonical quantum gravity models. Finally, in Section \ref{sec:7} we conclude and provide an outlook for future directions.

\section{Shared principles, different strategies}\label{sec:2}
Quantum field theory served as the starting point for early investigations into quantum gravity.
However, GR turned out to be non-renormalizable at  two-loops\cite{Goroff:1985th,Goroff:1985sz}. In view of the failure to apply standard perturbative field theoretical methods to GR, one could question whether the standard QFT approach could be revised and adapted appropriately for a gravitational system.

Compared to standard QFTs in flat spacetime, a QFT formulation for gravity does not only involve a given set of fields  and symmetries, corresponding to metric fluctuations and  diffeomorphism invariance respectively. A QFT formulation of gravity  has also to comply with the principle of \textit{background independence}.\cite{Kiefer:2005uk,Giulini:2006yg,Giulini:2006xi,Kiefer:2013jqa}  In fact, the realization of background independence can be considered as one of the most crucial challenges in a quantum field theoretic description of gravity and is what distinguishes quantum gravity from any other QFTs in curved spacetime.
Following the foundations of GR, none of the theory’s components, predictions, or assumptions should rely on a predetermined, fixed metric, but rather, the metric is a dynamical entity of its own.

In QFT, DeWitt's background field method is commonly used \cite{DeWitt:1964mxt,Abbott:1981ke,DeWitt:2003pm}. It involves  separating the metric $g_{\mu \nu}$ into a background  $\bar g_{\mu \nu}$ and a fluctuation $h_{\mu\nu}$. A possible separation is the linear splitting
\begin{equation}\label{eq:1}
g_{\mu \nu} = \bar g_{\mu \nu}+ h _{\mu \nu}\;.
\end{equation}
However, when applying this method it should be avoided  imposing a specific form on the background $\bar g_{\mu \nu}$, despite the appeal of certain well-behaved backgrounds that could simplify technical computations. In quantum gravity, it is essential  to leave the background  arbitrary and dynamical, requiring techniques that enable computations to be performed within such a general background framework. The choice of a background  has to be distinguished with the choice of a gauge in the case of gauge theories. In gauge theories, gravity included, one can freely choose a gauge, knowing that physical observables, such as scattering amplitudes, remain unaffected by this choice and thus render it inconsequential \cite{DeWitt:1980jv,Abbott:1980hw,Parker:2009uva}.

 In addition to choosing a gauge, in quantum gravity choosing a background represents a different conceptual step. The choice of a background does more than just provide a reference geometry: it effectively selects a specific momentum space structure and fixes the underlying spacetime topology. In fact, it determines the spectrum of differential operators (such as the d'Alembertian or Dirac operator), which in turn governs the behavior of propagators, functional traces, and flow equations. As a result, the spectral properties of the background can have a significant impact on the outcome of the entire computation. 
Furthermore, different backgrounds exhibit different isometry groups and thus support different numbers and types of symmetries. These symmetries affect not only the form of the action and allowed counterterms, but also constrain the dynamics and simplify computations through symmetry reduction. Consequently, results obtained in one background may not straightforwardly generalize to others, especially when trying to draw physically meaningful, background-independent conclusions.

 In principle, background independence can be  restored by summing the entire perturbation series in the fluctuation $h_{\mu \nu}$. One such attempt to deal with this issue is to  summing up the entire
 perturbation series
by calculating the flow of the tower of the $n$-point functions of the fluctuation \cite{Reuter:2019byg, Pawlowski:2020qer}. Alternatively, one
may try to solve simultaneously the Ward identity and the flow
equation \cite{Pawlowski:2005xe,Becker:2014qya}. Both of these routes have proven  to be  extremely challenging tasks.

Modern approaches to quantum gravity satisfy the requirement of background independence in two fundamentally distinct ways and can be divided into two classes \cite{Ashtekar:2014kba}:
\begin{enumerate}
\item \underline{Avoid using a gravitational background in any form}. These approaches aim to define the theory and explore its implications without relying on a background metric. This principle is followed for instance in Loop Quantum Gravity \cite{Thiemann:2007pyv}  and discrete methods in quantum gravity \cite{Loll:2019rdj,Ambjorn:2024pyv,Loll:2025eks}, which are based on different dynamical constituents, such as lengths or areas \cite{Dittrich:2021kzs,Borissova:2023yxs}. The background is then rather a structure emerging after the dynamics has been solved.
\item \underline{Keeping the   background metric arbitrary during the intermediate steps of} \underline{quantization}. This ensures that no physical predictions depend on the choice of metric at the final stage. The correct use of the background field method is central to the continuum-based gravitational  RG approach, which we will utilize for instance in the exploration of Asymptotic Safety. Crucially in this procedure, specifying no background is equivalent to considering all possible backgrounds \cite{Reuter:2008wj, Reuter:2019byg}. Moreover, after having solved the renormalized path integral, one can self-consistently establish the background's dynamic by imposing the quantum equations of motion \cite{Pagani:2019vfm,Reuter:2019byg, Ferrero:2022hor}.
\end{enumerate}

Next to background independence, in the efforts to overcome the perturbative renormalizability barrier identified by Goroff and Sagnotti \cite{Goroff:1985sz, Goroff:1985th},  it has been suggested that understanding gravitational interactions at the microscopic scale requires a shift toward \textit{non-perturbative methods} \cite{Ashtekar:1991hf,Thiemann:1996ay,Ambjorn:2000dv,Hamber:2004ew,Ashtekar:2004eh,Kiefer:2007ria,Ambjorn:2012jv, Reuter:2019byg}. This has been shared by approaches belonging to both classes.
Non-perturbative methods can capture effects that can not be captured by expanding physical quantities as a power series in a small coupling constant or parameter. Perturbative techniques approximate solutions by treating interactions as small corrections to free theories, valid only when the coupling is weak. Non-perturbative approaches, in contrast, deal directly with the full theory without assuming small parameters, allowing them to capture phenomena such as phase transitions, tunneling and strong coupling effects.

The realization of non-perturbativity varies significantly depending on the specific quantum gravity theory being considered and the mathematical framework employed.  For example, within a QFT framework such as  AS, non-perturbative flow equations are used \cite{Reuter:2019byg} and no perturbative expansion is needed. In a discrete framework, such as in the loop quantization of CQG, no truncation nor perturbative expansions are required. Rather, the quantum geometry is represented by means of spin networks, which are non-perturbative quantum states of space. They emerge from the non-perturbative, background-independent quantization of GR using the Ashtekar--Barbero variables and the holonomy-flux algebra \cite{Ashtekar:1991hf,Baez:1995md,Thiemann:1996aw,Thiemann:1996ay, Thiemann:2007pyv}.

In summary, background independence and non-perturbativity are two fundamental principles common to both Asymptotic Safety and Canonical Quantum Gravity, albeit implemented and realized in distinct ways \cite{Thiemann:2024vjx}.

{In addition to this, it should be noted that in this review both AS and CQG are formulated within a quantization paradigm, assuming a pre-existing classical spacetime structure to be quantized. In contrast, a number of other approaches take an emergent perspective, positing that the gravitational field is a collective phenomenon arising from more fundamental, possibly non-spatiotemporal, degrees of freedom. Examples include Group Field Theory (GFT) \cite{Oriti:2006se,Oriti:2011jm,Marchetti:2024nnk}, causal set theory \cite{Surya:2019ndm,Dowker:2024fwa}, and also, in a certain sense, canonical LQG and lattice-based formulations \cite{Ambjorn:2022naa}. In all these frameworks, the fundamental variables are not smooth fields but discrete or combinatorial structures, from which the continuum spacetime of GR can arise only through  collective dynamics \cite{Oriti:2018dsg}.}

{A related conceptual commonality concerns spacetime topology change. While recent continuum and discrete studies \cite{Giddings:1987cg,Horowitz:1990qb,Dowker:2002hm,Marolf:2020xie,Asante:2021zzh,Asante:2021phx,Asante:2025qbr}  have investigated quantum gravitational dynamics allowing for topology change,  both CQG and the AS-RG formulation usually assume fixed (globally hyperbolic) topologies. }

\section{Towards candidate observables for quantum gravity}\label{sec:3}
The desire to identify suitable observables that capture essential information about the theory of quantum gravity has  played an important role in the connection between Asymptotic Safety and Canonical Quantum Gravity.
 In this review by \textit{observable} we refer  to quantities which are diffeomorphism invariant (or gauge invariant). This makes them good candidates for physical detection \cite{Bergmann:1949zz,Bergmann:1954tc,Komar:1958ymq,Bergmann:1972ud,Bergmann:1960wb}.
\bigskip

To illuminate this connection we will provide in the following an overview of the two approaches.
CQG aims at  quantizing gravity based on the Hamiltonian formulation of GR. Spacetime is decomposed into a foliation of spatial hypersurfaces, evolving in time. This leads to a phase-space description where the fundamental variables are the three-dimensional spatial metric and its conjugate momentum \cite{PhysRev.116.1322}.  

A major objective in CQG is the identification of physical observables.  Unlike conventional QFTs, where observable information encoded in correlation functions or scattering amplitudes is defined with respect to a fixed background, the observables should not depend on the background.  It is  one of the principal goals of CQG,  to construct   fully background-independent observables.

With CQG  being formulated in the Hamiltonian framework, the evolution of spacetime is governed by constraints. These are the Hamiltonian constraint and the momentum constraint, which generate time evolution and spatial diffeomorphisms (see Ref. \citen{Henneaux:1994lbw} for a detailed analysis of Hamiltonian constrained systems).
Following the quantum mechanical prescription, a  physical observable is then an operator which commutes with all  constraints. This is the definition of a Dirac observable \cite{Dirac:1949cp, Dirac:1950pj, Dirac:1958sc,dirac1967lectures}. Such physical observables must be invariant under the gauge symmetries generated by the constraints, including both spatial and temporal coordinate transformations.

The physical observables form a closed algebra under Dirac brackets (the constrained system's generalization of Poisson brackets). While the constraints themselves may form a separate algebra, it is specifically the Dirac observables that maintain closure under this bracket structure. Within the process of quantization, these Dirac brackets translate into commutators. Furthermore, due to the constraint structure, the observables' algebra often exhibits non-trivial features, including possible central extensions or deformations \cite{Teitelboim:1972vw,Kuchar:1991qf,Kuchar:1993ne,Ashtekar:1994kv, Thiemann:2007pyv}.

In loop quantization, the fundamental variables are based on a reformulation of GR using the Ashtekar connection and the densitized triad. Instead of quantizing the connection and triad directly, the theory promotes holonomies, i.e., path-ordered exponentials of the Ashtekar connection along curves, and fluxes, i.e., surface integrals of the densitized triad, as the basic operators. Holonomies capture parallel transport and encode information about curvature, while fluxes measure geometric quantities like the area \cite{Ashtekar:1986yd,Ashtekar:1987gu,Rovelli:1989za,Ashtekar:1991hf,Ashtekar:1995zh}. Together, they form an algebra that encodes the quantum structure of spacetime. However, defining proper Dirac observables in terms of them remains an ongoing challenge. Unlike standard gauge theories, the Hamiltonian constraint in GR generates ``evolution'' in coordinate time, which is not physical in a background-independent theory. This leads to several problems, like the algebra of observables not being a Lie algebra \cite{DeWitt:1967uc, Kuchar:1980ht}, or making the algebra hard to represent quantum mechanically without anomalies (violations of the classical constraint algebra in the quantum theory) \cite{Thiemann:1996ay,Giesel:2006uj}, or regularization ambiguities \cite{Perez:2005fn,Liegener:2019dzj}. 
Furthermore,  defining observables requires solving the quantum constraints explicitly, especially the Hamiltonian constraint, something only partially under control, even in simplified models \cite{Thiemann:2007pyv}.

Since the Hamiltonian constraint generates time evolution, physical states $\psi$ must satisfy
\begin{equation}\label{H=0}
H \psi = 0\;.
\end{equation}
This equation suggests that the quantum state of the universe is frozen, leading to ambiguities in how to define evolution in the Schrödinger equation.
One possible resolution is the \textit{relational  approach} \cite{Rovelli:1989jn,Rovelli:1990pi,Rovelli:1990ph,Rovelli:2001bz,Dittrich:2005kc,Dittrich:2004cb,Tambornino:2011vg}. There, time is not an external parameter but a quantity defined with respect to other physical degrees of freedom present in the theory. For example, one may choose a physical variable, such as a matter field, as a ``clock" to describe physical time evolution. This procedure can be interpreted as a  ``physical gauge fixing". Another equivalent way to deal with the problem posed by \eqref{H=0} is the construction of observables using Dirac's algorithm \cite{Dirac:1949cp,Dirac:1950pj, Dirac:1958sc}. These two approaches are in general equivalent only prior to quantization \cite{Kunstatter:1991qe}: the gauge-fixing method eliminates from the phase space any unphysical degrees of freedom (\textit{reduced phase space}) \cite{Thiemann:2004wk}, while the construction à la Dirac does not. In many cases, only the first quantization approach can be successfully implemented, as the quantization of Dirac observables often proves to be highly challenging. 
\bigskip

On the other side, Asymptotic Safety is a QFT-based approach to quantum gravity that aims to ensure predictivity at all energy scales by identifying a suitable UV fixed point in the renormalization group  flow \cite{Weinberg:1976xy,Reuter:1996cp, Reuter:2019byg}. This fixed point guarantees that the theory remains well-defined at high energies, avoiding divergences that typically plague quantum gravity in standard perturbative QFT treatments \cite{Goroff:1985sz}.

One of the key advantages of AS is the possibility to systematically connect to the IR, where the observable, effective theory lives, by stepwise integrating out degrees of freedom. This is the modern view on renormalization introduced by Wilson \cite{Wilson:1973jj}.  The flow between UV and IR can be achieved in a robust way \cite{Codello:2015oqa,Ohta:2020bsc}. While the full quantum gravity theory is defined at the UV fixed point, at lower energies, one can transition to an effective description.  In the limit where all the degrees of freedom have been integrated out, following the running of the parameters under the renormalization group, physical observables can be  computed. In this limit, key quantities such as scattering amplitudes, correlation functions, and  observables can be accessed, providing a bridge between the fundamental (quantum gravity) theory and measurable predictions.

Importantly, in AS one must distinguish between the RG scale, which sets the coarse-graining resolution in the RG flow, and the physical momentum appearing in scattering processes or in the propagators. While the RG flow governs how effective couplings evolve under a change of the RG scale, this does not directly furnish  the running wrt. the momentum of external states.  It is the running with respect of the external momenta which represents the physical running \cite{Buccio:2023lzo,Buccio:2024hys,Buccio:2024omv}. How to connect these two runnings is a current active topic of investigation. A comparison between these two different runnings and their qualitative agreement in the graviton propagator can be found in Refs. \citen{Bonanno:2021squ, Fehre:2021eob}. Furthermore, recent work has suggested that the distinction between standard and physical running may depend on the choice of renormalization point, and that any running of the gravitational couplings might only manifest in curved spacetime, where curvature  sets a physical scale \cite{Kawai:2025wkp}.

Moreover, a key feature of the RG flow is the set of critical exponents, which describe how the couplings scale under the RG flow (RG scale running) in the vicinity of a fixed point. These exponents are universal \cite{Wegner:1974sla}, meaning they are independent of the specific details of the theory. Since they remain consistent across different formulations of the same model, they can be exploited to define the model. Furthermore, they determine the number of relevant, marginal, and irrelevant directions in the theory space. The  relevant directions are those corresponding to physically adjustable parameters; hence, having a finite number of them ensures predictivity. This universality of the critical exponents is central and is expected to play an important role in understanding the dynamics of quantum gravity and ensuring its predictive power. As such, critical exponents govern the scaling of couplings near fixed points and offer insight into the universal behavior of observables. However, in order to complete the connection to physically measurable quantities, a better understanding of the physical running is required.

\bigskip

The Lagrangian path integral approach in Asymptotic Safety encounters difficulties when it comes to defining observables in the context of gravity \cite{Baldazzi:2021fye}. The effective description it provides is typically not at the level of diffeomorphism-invariant observables. Furthermore, as a result of the  most commonly employed methods in the field, the obtained flow equations are affected by gauge dependence \cite{Gies:2015tca,Bonanno:2025tfj, DAngelo:2025yoy}. Recently some  more refined approaches have been introduced, which set up  flow equations for gauge-invariant operators \cite{Wetterich:2016ewc,Baldazzi:2021fye,Falls:2025tid}.

The challenge of defining gauge-invariant observables can be traced back to the fundamental issue of identifying well-defined observable quantities in classical gravity \cite{Torre:1993fq,Giddings:2005id,Tambornino:2011vg,Khavkine:2015fwa,Goeller:2022rsx,Giddings:2025xym}. Due to the diffeomorphism symmetry in gravity, observables must remain invariant under diffeomorphism transformations. This challenge has been solved within the relational formalism and has been particularly developed in Canonical Quantum Gravity \cite{Dittrich:2004cb,Thiemann:2004wk,Dittrich:2005kc,Dittrich:2006ee,Dittrich:2007jx, Dapor:2013hca}. In fact, it is the manifestation of the problem of time discussed earlier \cite{Kuchar:1991qf,Isham:1992ms}:  a well-defined observable can only be  specified relative to a chosen reference system or observer. The development of such an implementation within the Asymptotic Safety framework is a relatively recent advancement.

On the other hand, while CQG comes with a well-defined notion of observables, there is no fully established systematic mechanism for transitioning to an effective description in the  IR regime. Moreover, its notion of ``effective" dynamics  in general differs from the Lagrangian field theoretical one. 
Among the attempts to introduce coarse-graining in CQG-related approaches we mention coarse-graining in spin foams via tensor network renormalization \cite{Bahr:2009mc,Dittrich:2011zh,Bahr:2012qj,Dittrich:2013uqe,Dittrich:2014mxa,Bahr:2014qza} , also used to study perfect discretizations \cite{Dittrich:2008pw,Bahr:2011uj,Dittrich:2014ala}, and in Tensorial Group Field Theory (TGFT) \cite{BenGeloun:2016rqa,Dekhil:2024ssa}. {In particular, within TGFT,  both perturbative and FRG methods have been developed to explore the continuum and non-perturbative limits. The perturbative approach has been formulated directly in the language of spin-foam amplitudes, showing that these correspond to a specific truncation or sector of the complete TGFT dynamics \cite{BenGeloun:2011rc,BenGeloun:2013vwi}. The FRG formulation, on the other hand, provides a more general framework, allowing for the identification of fixed points and critical behavior \cite{Carrozza:2016tih,Pithis:2020sxm}. This makes possible, at least in principle, a systematic comparison of continuum-limit properties, such as critical exponents. In addition, mean-field and condensate methods have been applied to TGFTs, providing a technically simplified  way to probe continuum, collective regimes \cite{Gielen:2013naa,Oriti:2016qtz,Gerhardt:2018byq}.}

Furthermore, Hamiltonian renormalization methods have been recently advanced, focusing on constructing the correct continuum (ultraviolet) Hamiltonian via RG flows \cite{Lang:2017beo,Lang:2017yxi,Zarate:2025erv}. It is important to emphasize that these advancements are aimed at the construction of the continuum limit.  Remarkably, effective spin foams have been recently developed, which allow computational access to the continuum limit \cite{Dittrich:2022yoo,Borissova:2022clg} and only recently the obtained continuum limit  has been analyzed via  FRG methods \cite{Borissova:2025frj}. 

{Finally, a number of studies have explored the reconstruction of the graviton propagator from spin foam amplitudes \cite{Alesci:2007tx,Alesci:2007tg,Bianchi:2009ri,Bianchi:2011hp,ChaharsoughShirazi:2015uuu,Han:2017isy} and the access to a low-energy, perturbative regime through large-spin expansions \cite{Han:2013tap,Han:2018fmu,Han:2021kll}. While these investigations have provided valuable insights into how linearized gravitational correlations could, in principle, emerge from spin foam dynamics, it must be emphasized that they rely on extremely simple simplicial complexes and very limited degrees of freedom. As such, they probe only a small portion of the full theory’s dynamics and remain far from testing the continuum limit, which is essential for any meaningful notion of graviton propagation. In this respect, a substantial open challenge for the spin foam and canonical frameworks alike lies in demonstrating how continuum perturbative gravity can emerge from the nonperturbative discrete dynamics.}

While these developments have led to significant progress, it remains evident that, although Canonical Quantum Gravity offers a background-independent formulation of quantum gravity, it still lacks a fully established renormalization group framework for systematically deriving low-energy effective field theory descriptions in the sense understood within QFT (see next subsection). Developing a robust approach to extract effective IR physics from canonical methods remains a challenging problem. This makes it arduous to directly connect the fundamental quantum gravitational dynamics to observable predictions at macroscopic scales: bridging this gap might be essential for linking quantum gravity to experimentally accessible regimes.

\subsection{``Effective" physics: a semantic explanation}
It is important to emphasize that the term \textit{``effective''} typically refers to different concepts in the context of quantum field theory and canonical approaches \cite{Cametti:1999ii}. In the context of QFT, effective typically refers to a framework obtained by integrating out high-energy degrees of freedom, without changing the physical content of the theory. The effects of the eliminated degrees of freedom are captured, for instance, in the running of coupling constants  \cite{Wilson:1983xri,Polchinski:1983gv} or Wilson coefficients in effective field theory \cite{Georgi:1993mps,Burgess:2020tbq}.

Within LQG, an effective theory refers to three different procedures
\begin{enumerate}
\item   Semiclassical approximation obtained by taking expectation values of quantum operators on suitably chosen semiclassical states (e.g., coherent states). The result gives rise to an effective Hamiltonian (or equations of motion) that includes quantum corrections to classical dynamics.
\item Modified  equations of motion and Hamiltonian that approximate quantum dynamics, derived  by incorporating key quantum features such as holonomy corrections.
\item Reconstruction of an effective Lagrangian  to reproduce the quantum-corrected equations of motion, providing a classical-like action that encodes quantum gravity effects for studying phenomena like cosmological dynamics or black hole interiors. This effective Lagrangian is typically not unique.
\end{enumerate}
Within LQG, the effective dynamics resulting from procedure (1) and (2) has been  largely investigated \cite{Bojowald:2005cw,Bojowald:2006zi,Bojowald:2006ww,Bojowald:2007zz,Bojowald:2007nho}, especially for black holes \cite{Bojowald:2018xxu,Han:2022rsx,Giesel:2023hys,Belfaqih:2024vfk,Zhang:2024khj} and cosmology \cite{Bojowald:2012xy,Langlois:2017hdf,Dapor:2017rwv,Dapor:2019mil}. However, the precise correspondence with the covariant effective theory remains an open question and is still under investigation
\cite{Bojowald:2024beb}.

In particular, within the procedure sketched in (3), several works have shown that the non-perturbative quantum dynamics of background-independent approaches (especially in Loop Quantum Cosmology (LQC) \cite{Ashtekar:2011ni,Li:2023dwy}  and GFT \cite{Oriti:2006se,Marchetti:2024nnk}) can be recast as classical modified-gravity actions. In exceptional cases one inverts the effective equations from the quantum Hamiltonian (Legendre transform) to find a covariant action. In others a suitable modified Lagrangian is postulated whose field equations reproduce the same dynamics as the effective Hamiltonian. As an example,  a unique Palatini $f(R)$ action whose field equations exactly reproduce the isotropic LQC bounce has been derived \cite{Olmo:2008nf}. Starting from the LQC effective Friedmann equation for a massless scalar, one can solve for a connection-independent Lagrangian $f(R)$  that yields the same dynamics. The resulting covariant action (an infinite series in $R$) leads to exactly the non-singular bounce of LQC.
Subsequently, it was shown that metric $f(R)$ and higher-order gravity theories can reproduce LQC dynamics, treating higher-derivative terms as perturbative corrections around GR \cite{Sotiriou:2008ya}.  Later works extended this method to more general curvature invariants and modified LQC models, deriving families of higher curvature actions that mimic the LQC bounce phenomenologically \cite{Ribeiro:2021gds}. Independently, mimetic and degenerate higher-order scalar–tensor theories, particularly the Chamseddine--Mukhanov’s limiting-curvature model \cite{Chamseddine:2013kea}, were shown to exactly reproduce LQC’s bounce equations in isotropic settings, offering an alternative and more precise  correspondence \cite{Langlois:2017hdf,Han:2022rsx}.

{Always within (3), among LQG covariant path-integral quantum gravity models, spin foam models offer a path integral formulation of quantum gravity \cite{Rovelli:2014ssa,Engle:2023qsu,Livine:2024hhc}, from which one can extract an effective, semiclassical Lagrangian. It has been shown that, if the vertex amplitude exhibits the correct large-spin behavior, the spin-foam path integral reproduces Regge calculus in this discrete semiclassical limit \cite{Conrady:2008mk,Barrett:2009gg,Mikovic:2011zx}. While this represents an important consistency check, it should be emphasized that reproducing classical Regge calculus on a fixed simplicial complex is not equivalent to recovering continuum GR. The correspondence concerns the discrete classical regime, not the continuum limit, and several open issues remain even at the level of the large-$j$ approximation \cite{Bonzom:2009hw,Bahr:2014qza,Dittrich:2014ala,Asante:2020iwm}. Indeed, as illustrated by the experience with simplicial quantum gravity approaches such as quantum Regge calculus and causal or dynamical triangulations \cite{Hamber:2009mt,Ambjorn:2012jv}, establishing a genuine continuum limit from a discrete quantum theory poses substantial conceptual and technical challenges, despite the fact that discrete models trivially reproduce classical Regge calculus by construction.}

It needs to be emphasized, that also within AS a procedure similar to (3) can be performed. This goes  under the name of \textit{RG improvement}. Interpreting the running effective  action as af a scale-dependent effective Lagrangian that includes quantum corrections, the quantum-corrected equations of motion can be derived. 
The process of RG improvement consists then in evaluating couplings at a dynamically chosen scale (e.g. the radius of a black hole \cite{Bonanno:1998ye,Bonanno:2000ep,Koch:2014cqa,Eichhorn:2022bgu,Platania:2023srt} or the cosmic time \cite{Bonanno:2001xi,Bonanno:2002zb,Bonanno:2017pkg}).
This procedure fits naturally into the idea of using an effective Lagrangian to study semiclassical or phenomenological consequences, despite the underlying background dependence and ambiguity in scale setting. 

{This comparison also highlights a deeper conceptual distinction between the two frameworks. In LQG, coarse-graining does not simply correspond to a cutoff in modes of the same kind, as in the AS  RG-based approach, but rather to a transition between qualitatively different kinds of microscopic structures. Even if the discrete geometries underlying LQG were regarded as regularization devices, the reconstruction of a smooth spacetime would still amount to an emergent, collective phenomenon.}

{Related to this point, it is important to distinguish clearly between semiclassical and continuum approximations or limits. The former concern the recovery of classical dynamics from quantum operators on suitable semiclassical states, while the latter refer to the emergence of smooth spacetime behavior from the collective dynamics of fundamentally discrete or combinatorial structures. These two notions are conceptually and technically distinct, and the continuum limit, rather than the semiclassical one, is what captures the genuine coarse-grained, collective dynamics discussed above. Clarifying this distinction is essential when comparing the effective descriptions arising in approaches such as LQG, spin foam models, or GFT with those obtained through the functional RG in AS.}

\bigskip
In the following when we will refer to ``effective" we will mean the QFT-based meaning introduced at the beginning of this subsection, seeking to  connect it with the CQG formalism.

\bigskip
For the reasons introduced in this section, in the attempt of making observable predictions in a theory of quantum gravity, there are ongoing efforts to merge the two approaches, each addressing the challenges faced by the other. Before we introduce the approach to fruitfully combine them, it is important to present which are the obstacles in relating the covariant and the canonical quantum gravity formalism.

\subsection{Gravitational Hamiltonian and gravitational Lagrangian}
It is a longstanding open issue, how  the Lagrangian and the Hamiltonian formalism can be related in a gravitational context (see appendix E.2 in Ref. \citen{Wald:1984rg}). This problem is indeed rooted in the need of defining a time in the Hamiltonian context. 

Consider for instance the Einstein--Hilbert action in $d$ dimensions (for simplicity we consider for the moment $\Lambda = 0$)
\begin{equation}
S= \frac{1}{16 \pi G_N}\int \text{d}^dx \sqrt{-g}R[g]\;.
\end{equation}
To move forward to the Hamiltonian formulation we foliate the spacetime into spatial slices labeled by a time parameter $t$
\begin{equation}\label{gmunu}
\text{d}s^2= -N^2\text{d}t^2 + q_{ij} (\text{d}x^i+N^i \text{d}t)(\text{d}x^j+N^j \text{d}t)\;,
\end{equation}
where $N$ is the lapse function, $N^i$ the shift vector, and $q_{ij}$ the $(d-1)$ spatial metric on constant time slices. The action becomes
\begin{equation}
S=\frac{1}{16\pi G} \int \text{d}t \text{d}^{d-1}x N \sqrt{q} \left({}^{(d-1)}R[q] + K_{ij}K^{ij}-K^2\right)\;,
\end{equation}
where ${}^{(d-1)}R[q]$ is  the $(d-1)$-dimensional curvature constructed from $q_{ij}$ and $K_{ij}$  the extrinsic curvature
\begin{equation}
K_{ij} = \frac{1}{2N} \left(\dot q_{ij} -D_iN_j -D_j N_i\right)\;.
\end{equation}
 Here $D$ is the spatial covariant derivative compatible with $q_{ij}$. We can proceed in defining the canonical momenta. The momentum conjugate to $q_{ij}$ is
\begin{equation}
\pi^{ij} = \frac{\delta S}{\delta \dot q_{ij}} = \sqrt{q} \left(K^{ij}-q^{ij}K\right)\;,
\end{equation}
while the momenta related to $N$ and $N^i$ both vanish. These represent primary first-class constraints and are hence Lagrange multipliers.

The Hamiltonian is then defined as
\begin{equation}\label{H}
H = \int \text{d}^{d-1}x \left(\pi^{ij}\dot q_{ij}- \mathcal{L}\right) = \int \text{d}^{d-1}x \left(N \mathcal{H}+ N^i \mathcal{H}_i\right)\;,
\end{equation}
where we defined the Hamiltonian constraint density
\begin{equation}
\mathcal{H} = \frac{1}{\sqrt{q}} \left(\pi^{ij} \pi_{ij}- \frac{1}{2}\pi^2\right)-\sqrt{q}{}^{(d-1)}R 
\end{equation}
and the diffeomorphism (momentum) constraint density
\begin{equation}
\mathcal{H}_i = -2 D_j \pi^j_i\;.
\end{equation}
Therefore, by varying \eqref{H}  with respect to $N$ and
$N^i$, the dynamics is entirely governed by the constraints
\begin{equation}
\mathcal{H}\approx 0\;, \qquad \mathcal{H}_i\approx 0\;.
\end{equation}
Here, the symbol  $\approx$ is called ``weakly equal to'' which simply means that this equality needs to
be satisfied only on the constraint hypersurface in phase space.
As it turns out,  the Hamiltonian $H$ is a sum of constraints and vanishes on-shell. This is characteristic of reparametrization-invariant theories like GR, where there is no absolute time evolution. 

As mentioned above, in order to define the dynamics  one can resort to the relational formalism. However, the relational procedure necessarily selects a preferred time, which breaks the covariance upon which the Lagrangian is based (see Section \ref{sec:5}).

While the relational formalism introduces dynamics by selecting a preferred temporal reference, the Lagrangian approach preserves full covariance and instead handles the redundancy associated with gauge symmetries through a gauge-fixing procedure and the introduction of ghosts à la Faddeev--Popov \cite{Faddeev:1967fc,faddeev1969feynman} is needed, in order to account for the unphysical degrees of freedom.
In this context, gauge-fixing independence has been proven to hold only in a perturbative loop expansion, imposing order by order on-shell conditions \cite{Parker:2009uva}. In this setting, gauge-invariant observables are obtained from evaluating correlation functions or amplitudes on-shell. However their construction typically involves the choice of a background. Conversely, in non-perturbative settings such as in AS, the results are still affected by gauge-parameter dependence \cite{Benedetti:2010nr,Gies:2015tca,Bonanno:2025tfj,DAngelo:2025yoy}.  Furthermore, the regularization method chosen also generally  affects the full quantum gauge invariance \cite{Reuter:2019byg}. In order to alleviate this problem, within AS the form factor program  has been established \cite{Knorr:2019atm,Draper:2020bop,Draper:2020knh,Knorr:2021iwv,Knorr:2022dsx}: the idea is to study the non-local effective action for gravity (and matter fields) using form factors, which are Laplacian- or momentum-dependent functions appearing in front of curvature invariants (or other operators). Form factors encode the scale-dependent (and generally non-local) structure of quantum corrections. This representation allows one to keep background gauge invariance manifest throughout the RG flow, which contrasts with ansätze based on local operators in the Einstein--Hilbert or polynomial expansions that often introduce artificial gauge or parametrization dependence. Because form factors resemble quantities that appear in on-shell amplitudes, they bring the framework closer to settings where gauge invariance is manifest.

Although canonical constraints and the covariant gauge-fixing procedure have been related in special cases,  it is difficult to transition from one to the other in general.
This difference arises due to different concepts of gauge symmetry in the Lagrangian and Hamiltonian formalism. In the Lagrangian approach, the Lagrangian dictates the gauge group; within this group, the gauge transformations are purely kinematical and are insensitive to the exact form of the Lagrangian compatible with that given gauge group. In fact, they follow from a Lie algebra (for example, in GR, this is the Lie algebra of vector fields, related to diffeomorphisms).
In the Hamiltonian framework, gauge transformations are generated by constraints that do not form a conventional Lie algebra in GR. This is because their structure functions depend on phase-space variables rather than being constant. Consequently, the transformations generate an open algebra (a Lie algebroid) characterized by non-constant structure functions. These gauge transformations collectively constitute the Bergmann--Komar group, which encodes the full diffeomorphism symmetry of GR in a phase-space dependent manner \cite{Bergmann:1972ud, Bergmann:1981fc}. The Bergmann--Komar group essentially determines the dynamical symmetry of  GR, namely the symmetry that relates entire families of solutions of the equations of motion and depends on the specific form of the action or Hamiltonian.

When fields satisfy classical equations of motion (e.g., Einstein’s equations in GR), phase-space variables are no longer independent. This reduces structure functions to structure constants (numerical coefficients), closing the algebra into a conventional Lie algebra. Consider for instance the constraint algebra in GR
\begin{equation}
\{\mathcal{H}(x), \mathcal{H}(y)\} \propto q^{ij} \delta_{,j} (x,y) \mathcal{H}_{i}(x)\;.
\end{equation}
Off-shell $q^{ij}$ is arbitrary, making the algebra open. On-shell instead it becomes tied to the physical reference fields that deparametrize: the structure functions $q^{ij}$ collapse to fixed values yielding structure constants.
As a result, in gauge systems the Hamiltonian and covariant path integral formulations of gauge symmetry coincide only on-shell. However, the path integral inherently operates off-shell, integrating over all field configurations regardless of whether they obey classical dynamics. Since classical solutions constitute a measure-zero subset in this infinite-dimensional space, the quantum theory cannot rely on on-shell equivalence to resolve gauge redundancies. In this context, so called on-shell approaches \cite{Benedetti:2010nr,Falls:2024noj}, may help alleviate this discrepancy, though this remains to be  established. Nonetheless, this mismatch complicates the quantization of gravity, where gauge invariance is tied to diffeomorphisms, and highlights the need to connect gauge-fixing procedures to physical relational observables.\footnote{In a series of works it has been  clarified how gauge fixing, observables and the Lagrangian versus Hamiltonian formalisms interplay in generally covariant theories, especially in GR. Demanding equivalence between the Lagrangian and Hamiltonian descriptions forces the true gauge generators to depend on the lapse and shift function \cite{Pons:1996av,Pons:2009cz,Pons:2010ad}.}

{It should be noted at this stage, that the introduction of a ``preferred" reference in the relational strategy refers to a physical frame, typically implemented through dynamical reference fields, and not to a coordinate choice or gauge-fixing. When formulated in this way, the relational construction can be implemented entirely at the Lagrangian level while maintaining full diffeomorphism invariance of the resulting observables, as exemplified in generally covariant relational frameworks (non-reduced phase space) and in applications such as TGFT cosmology \cite{Gielen:2013naa,Oriti:2016qtz,Marchetti:2024nnk}. This was also the path followed in the construction of Ref. \citen{Baldazzi:2021fye}.
In principle, therefore, the relational formalism is perfectly compatible with general covariance: the introduction of reference fields merely provides a way to parametrize physical observables in a diffeomorphism-invariant theory. What changes is the representation of dynamics, expressed relationally with respect to the chosen physical frame, rather than the symmetry content of the theory.
In practice, however, concrete implementations often involve approximations that single out a particular physical frame or neglect the backreaction and quantum properties of the reference system. In such cases, covariance may appear broken or only approximately realized, even though the underlying formalism remains covariant. The challenge is thus not a conceptual incompatibility with general covariance, but the technical problem of embedding relationally defined, on-shell observables into an off-shell covariant quantization framework.}

Summarizing, the equivalence between the Hamiltonian and  Lagrangian formulations can in general not be established because of the following reasons:
\begin{itemize}
\item Non-commutativity of reduction and quantization process: the order of reduction (solving constraints) and quantization affects  the resulting quantum theory; that is, quantizing the reduced phase space does not yield the same physics as reducing after quantization.
\item  Inequivalence of gauge fixing and constraints: The Faddeev--Popov gauge-fixing procedure is not  equivalent to imposing  first-class constraints in the Hamiltonian formalism.
\item  Gauge dependence of observables: Physical observables are not  invariant under different choices of Faddeev--Popov gauge fixing, especially in covariant non-perturbative regimes and in the context of gravitational theories.
 \item   Impossibility to reconstruct the covariant dynamics: Given a choice of preferred time direction in the Hamiltonian framework, it is not always  possible to reconstruct a covariant Lagrangian description of the quantum dynamics.
\end{itemize} 
Hence, a more physical connection between the formalisms is needed: This will be achieved by means of the explicit introduction of the relational observables in the path integral and  will be the subject of Section \ref{sec:6}.

A closely related issue which has been  source of  concern is the choice of the measure. While typically from a Lagrangian point of view, the measure is not uniquely fixed, provided it satisfies the given symmetry conditions, from a Hamiltonian (relational) perspective, the construction of the physical configuration space and its conjugate momentum space depends on the choice of clock and is hence fixed.
This might give rise to differences in local measure factors, discussed since the 1960s in geometrodynamics and background-dependent quantizations \cite{Leutwyler:1964wn,Fradkin:1973wke,Fradkin:1977hw,Gitman:1990qh,Branchina:2024lai}. These were argued to  contribute only to divergent terms in higher-loop amplitudes and to be dependent on the chosen regularization scheme. While certain regularization choices can cancel or suppress these contributions, the ambiguity surrounding their role has been recently been clarified \cite{Bonanno:2025xdg}. These  factors are however essential, as they properly reflect the off-shell symmetries generated by the constraints \cite{Bergmann:1981fc,Han:2009bb}.
Importantly, when relating the Hamiltonian framework to the Lagrangian path integral, the measure derived from the conjugate variables is also fixed and it may vary depending on the chosen physical frame and might be highly non-trivial. We will come back to this in Section \ref{sec:6}. 


\subsection{Remarks  on more empirical connections between gravitational Hamiltonian and Lagrangian}
The relationship between the gravitational Lagrangian and Hamiltonian has also been explored at various more advanced levels, including  the development of covariant Loop Quantum Gravity (spin foam models), in symmetry-reduced settings, and through the covariantization of effective equations of motion.

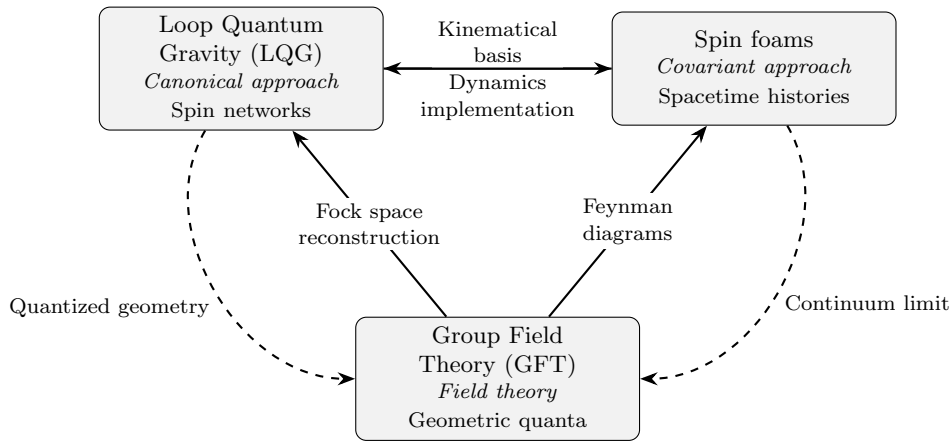
\begin{figure}
	\centering
	\begin{tikzpicture}[
		node distance = 3cm,
		concept/.style = {
			rectangle,
			draw,
			rounded corners,
			fill=gray!10,
			text width=3.5cm,
			minimum height=1.5cm,
			align=center,
			font=\small
		},
		arrow/.style = {
			->,
			>=Stealth,
			thick
		},
		label/.style = {
			font=\footnotesize,
			midway,
			fill=white,
			inner sep=2pt
		}
		]
		
		\node[concept] (lqg) {Loop Quantum Gravity (LQG) \\ \footnotesize \textit{Canonical approach} \\ Spin networks};
		\node[concept, right=of lqg] (spinfoam) {Spin foams \\ \footnotesize \textit{Covariant approach} \\ Spacetime histories};
		\node[concept, below=2.5cm of $(lqg.south)!0.5!(spinfoam.south)$] (gft) {Group Field Theory (GFT) \\ \footnotesize \textit{Field theory} \\ Geometric quanta};
		
		\draw[arrow] (lqg) -- node[align=center,label, above]{Kinematical\\basis} (spinfoam);
		\draw[arrow] (spinfoam) -- node[align=center,label, below] {Dynamics\\implementation} (lqg);
		\draw[arrow] (gft) -- node[align=center,label] {Fock space\\reconstruction} (lqg);
		\draw[arrow] (gft) -- node[align=left,label] {Feynman \\diagrams} (spinfoam);
		\draw[arrow, dashed] (lqg) to[out=-120,in=180] node[label, below left] {Quantized geometry} (gft);
		\draw[arrow, dashed] (spinfoam) to[out=-60,in=0] node[label, below right] {Continuum limit} (gft);

	\end{tikzpicture}
	\caption{LQG, Spin Foams, and GFT offer complementary perspectives on CQG. They are connected through a coherent mathematical structure. LQG is formulated in the canonical framework, where quantum states of geometry are described by spin networks, graphs labeled by SU(2) representations that encode quantized areas and volumes. These spin networks form the kinematical basis for spin foam models, which provide a covariant, path-integral approach by assigning transition amplitudes to histories of spin networks, represented as 2-complexes (foams). Spin foams thus aim to implement the dynamics of LQG in a spacetime setting. GFT, in turn, offers a field-theoretic generalization of this picture: it is a quantum field theory on group manifolds whose Fock space recreates spin network states as many-body excitations of geometric quanta. Perturbative expansions of GFT yield Feynman diagrams that correspond  to spin foam amplitudes, allowing the theory to generate sums over discrete spacetime geometries. Through coarse-graining, GFT also provides a pathway toward recovering the continuum limit, unifying the discrete structures of LQG and spin foams within a second-quantized framework. }\label{fig:111}
\end{figure}

Let us briefly comment on the relation between canonical and covariant LQG \cite{Rovelli:2014ssa,Engle:2023qsu,Livine:2024hhc} (see Figure \ref{fig:111} for a schematic connection, including GFT \cite{Livine:2011yb}). The covariant spin foam path integral represents the sum over histories of spin networks in the framework of LQG. In other words, it can be viewed as the projector onto solutions of the Hamiltonian constraint of LQG.\cite{Reisenberger:1996pu,Rovelli:1998dx}  In this picture the spin foam sum builds a Lagrangian and  emerges from “group-averaging” the Hamiltonian operator. However, this connection seems to be only formal: one writes the physical inner product as an (infinite) product or exponential of the Hamiltonian constraint.  Motivated by the idea that spin foam models can be viewed as expansions of the projector onto solutions of the Hamiltonian constraint operator,  a new regularization of Thiemann’s Euclidean Hamiltonian constraint was found, whose action matches spin foam dynamics \cite{Alesci:2010gb}.
In spite of that,  the spin foam ansatz is not directly obtained from the Hamiltonian, and this derivation remains incomplete, missing a  bidirectional link between canonical Hamiltonian and  spin foam Lagrangian. Consequently, to date, the precise relationship between spin foam models (such as the Engle--Pereira--Rovelli--Livine (EPRL) \cite{Engle:2007uq,Engle:2007wy,Engle:2007qf}) and canonical quantization of the Hamiltonian constraint remains an open question \cite{Thiemann:2013lka, Alesci:2013kpa,Han:2020chr}.

{It is also important to note that current spin foam models and their GFT formulations differ structurally from canonical LQG in a crucial way. The covariant models are rooted in simplicial geometry, employing piecewise-flat building blocks already in the definition of the quantum amplitudes \cite{Barrett:1997gw,Reisenberger:1996pu,Engle:2007wy,Perez:2012wv,Rovelli:2014ssa}. This simplicial setting restricts the underlying graphs or complexes to fixed valence and makes it difficult to establish a clear correspondence with the canonical quantum dynamics, in particular with a Hamiltonian constraint formulated for smooth geometries \cite{Dittrich:2007wm,Dittrich:2011vz}. Although various generalizations at the combinatorial level have been proposed to bring spin foam and GFT frameworks closer to the canonical theory \cite{Dittrich:2011vz,Oriti:2014yla,Gielen:2016dss}, their amplitudes remain tied to the simplicial construction and lose their geometric justification when extended beyond it. This mismatch is reflected mathematically in the failure of cylindrical consistency, a key feature of the kinematical Hilbert space in canonical LQG, which is not satisfied by any of the current spin foam or GFT models \cite{Bonzom:2011hm,Bahr:2014qza,Dittrich:2014ala,Asante:2020iwm}.}

Furthermore, one might analyze the correspondence on the level of symmetry-reduced models. Constructing the semiclassical or coarse-grained form of the Hamiltonian constraint operator typically involves taking expectation values in suitable states, particularly when considering  symmetry‐reduced models. For example, in LQC  one obtains effective Hamiltonians by coherent‐state or WKB methods and these match results from covariant path‐integral approaches. It was shown explicitly that in flat Friedmann--Lemaître--Robertson--Walker (FLRW) spacetimes, LQC with a massless scalar a group‐averaged path integral (spin foam analogue) yields exactly the same effective Hamiltonian constraint as the canonical theory \cite{Huang:2011es}. Crucially, several results are obtained assuming homogeneity and isotropy. This simplifies the quantum geometry to a finite number of degrees of freedom, allowing for analytical solutions in symmetry-reduced models. The use of a scalar field as an internal clock, combined with group-averaging to impose the Hamiltonian constraint, enables the construction of physical states and extraction of dynamics within this framework.

Beyond minisuperspace, one can derive effective Hamiltonians from full LQG using semiclassical (coherent‐state) techniques \cite{Thiemann:2000bw,Thiemann:2000ca,Han:2019vpw,Han:2020chr,Han:2020iwk}. Importantly, coherent state path integral techniques have been used to compute the one-loop effective action \cite{Ferrero:2025est}, representing the first computation of the one-loop effective action  of a loop-quantized model.  Moreover, the expectation value of Thiemann’s Hamiltonian operator on coherent states peaked on flat FRW data has been evaluated \cite{Dapor:2017rwv}. It turned out that the effective Hamiltonian constraint  for cosmology  includes corrections from the full theory, notably a Lorentzian term absent in simpler LQC schemes.  Hence, comparing the effective Hamiltonian to LQC provides a consistency check and links canonical LQG to emergent cosmological Lagrangians \cite{Alesci:2014uha, Han:2019vpw}.

\bigskip
Having discussed attempts and open challenges towards the connection of the Hamiltonian and the Lagrangian formalism at a classical and a quantum level, let us now give a closer look to the Lagrangian and the Hamiltonian approaches in quantum gravity, emphasizing their strengths and weaknesses.

\section{Asymptotic safety: surprises from a background-independent QFT}\label{sec:4}

Standard QFT approaches applied to investigate  the renormalizability of GR, starting by the early work  by 't Hooft and Veltman \cite{tHooft:1974toh} and Deser \cite{Deser:1974cy} also including matter, by Christensen and Duff including a cosmological constant \cite{Christensen:1979iy} and by Goroff and Sagnotti \cite{Goroff:1985sz, Goroff:1985th} and van de Ven \cite{vandeVen:1991gw} at two loops, typically apply Lagrangian perturbative renormalization  methods in flat spacetime background. They use dimensional regularization within the minimal subtraction scheme. This results in what is commonly referred to as the ``problem of renormalizability of quantum gravity": in GR, gravitational interactions are non-renormalizable at two loops, as they require an infinite tower of counterterms to cancel divergences.

In recent years, significant progress has been made in understanding  what is special about gravity and in adapting quantum field theory methods to accommodate gravitational interactions. As proposed by Weinberg in 1979 \cite{Weinberg:1980gg}, even slight modifications to standard methods can drastically change the results on renormalizability. These modifications we will introduce along this section turned out to be  essential to respect foundational principles of general relativity, such as background independence and diffeomorphism invariance.  Furthermore, it was suggested that dimensional regularization has to be revisited, as the regularization on the dimension $g_\mu^\mu = d$  is a regularization on the trace  of a dynamical field, namely the metric field \cite{Martini:2021slj,Martini:2022sll,Falls:2024noj}. This scenario was particularly studied in $d = 2+\epsilon$ dimensions \cite{Kawai:1989yh,Jack:1990ey}. Weinberg proposed that under appropriate modifications, gravity might be \textit{asymptotically safe}: gravitational scattering amplitudes could be finite even at scales beyond the Planck scale. This behavior may reflect the existence of a non-trivial UV fixed point in the RG flow, realized through an intrinsically non-perturbative mechanism.

Wilson's perspective on renormalization was equally revolutionary \cite{Wilson:1973jj,Wilson:1974mb,Wilson:1975hz,Wilson:1983xri,Kadanoff:2000xz}. It introduced the concept of coarse-graining, or averaging over short-distance fluctuations, with the resulting description captured by the effective action. This approach also introduced a fundamentally non-perturbative viewpoint and methodology. In particular, this setup was initially applied and successfully tested  in the study of critical phenomena of low energy systems \cite{Wilson:1983xri}.

\subsection{The Functional Renormalization Group}

Wilsonian-inspired methods, combined with the requirement to uphold background independence, converged into a powerful tool: the Functional Renormalization Group.
This method introduces a scale-dependent effective action, the Effective Average Action (EAA), which interpolates between bare and full-fledged effective action \cite{Wetterich:1992yh,Reuter:1992uk,Reuter:1993kw}. Remarkably, this action satisfies an exact, non-perturbative renormalization group equation \cite{Wetterich:1992yh,Morris:1993qb}.

The central idea of the EAA is to produce a  modification of the  partition function by integrating out high momentum modes and simultaneously suppress low momentum modes.
For instance, for a real scalar field $\hat \phi$, on Euclidean spacetime\footnote{We will comment later on the differences between Euclidean and Lorentzian signature.} $\mathbb{R}^d$ this is achieved by the addition of a scale-dependent  IR “cutoff functional” $\Delta S_k[\hat \phi]$ in the exponent of the integrand, leading to the definition of the modified partition function
\begin{equation}
	Z_k[J] = \int\mathcal{D}  \hat\phi \;\text{exp} \left(-S[\hat\phi]-\Delta S_k[\hat\phi]+ \int \text{d}^dx J(x)\hat\phi(x)\right)\;.\label{eq:Z}
\end{equation}
Here $J$ is an external source, and $S$ denotes the  bare action., which is left generic at this point.
The mode suppression term introduces  a momentum scale $k$,
\begin{equation}
	\Delta S_k[\hat\phi]= \frac{1}{2} \int\text{d}^dx\; \hat\phi(x)\mathcal{R}_k(\Box)\hat\phi(x)\;.
	\label{eq:DeltaS}
\end{equation}
By construction, the cutoff kernel $\mathcal{R}_k$ acts  as an infrared cutoff. For this purpose it has to have a functional form in the eigenvalues of the d'Alembertian $\Box$, such that it leaves the high-momentum modes unaffected, i.e., they will be integrated out in the partition function, while it generates   a mass-like contribution for the IR modes. Moreover, the cutoff kernel 
is designed to vanish as $k = 0$, ensuring the recovery of the full quantum effective action, and to diverge as $k \to \infty$,  freezing all fluctuations at the ultraviolet scale.

The partition function $Z_k$ associated to the one parameter family of bare actions $S+\Delta S_k$ gives rise to  the generating functional $W_k[J] = \ln Z_k[J]$. 
The next steps toward the definition of the EAA are similar to the standard textbook procedure which leads to the effective action. One defines the ($k$-dependent) field expectation value $\phi = \langle\hat \phi\rangle = \delta W_k/\delta J$ and computes the Legendre transform of $W_k[J]$ with respect to $J(x)$, at fixed $k$. If the functional relationship $\phi=\phi[J]$ can be solved for the source to yield $J = \mathcal{J}_k [\phi]$, it takes the form
\begin{equation}
	\tilde \Gamma_k[\phi] =\int \text{d}^dx\; J[\phi](x)\phi(x)+ W_k[J_k[\phi]]\;.
\end{equation}
Finally, subtracting the cutoff term $\Delta S_k$ we arrive at the definition of the EAA $\Gamma_k$:
\begin{equation}\label{eq:EAA}
	\Gamma_k[\phi] =	\tilde \Gamma_k[\phi]-\Delta S_k[\phi]=\int \text{d}^dx\; J[\phi](x)\phi(x)+ W_k[\phi]- \Delta S_k[\phi]\;.
\end{equation}

The EAA is closely related to a generating functional for correlators of field averages, hence its name.
In fact, from the EAA  it is possible to directly compute  all the ($k$-dependent) Green’s functions of the quantum theory by functional differentiation of $\Gamma_k$, without  any additional functional integration.
Moreover, it represents the scale-dependent  version of the standard effective action $\Gamma$.   As it has been defined and  given the required properties of the regulator, in the $k \to 0$ limit it approaches the  standard effective action $\Gamma_{k\to 0} \to \Gamma$, while in the UV limit $k \to \infty$, it approaches the bare action $\Gamma_{k\to \infty} \to S$.

 Importantly the EAA   satisfies the  exact Functional Renormalization Group Equation (FRGE) or Wetterich equation
\begin{equation}
	k	\partial_k	\Gamma_k[\phi] = \frac{1}{2}\;\text{Tr}\;\left[ (\Gamma_k^{(2)}+\mathcal{R}_k)^{-1}\;k\partial_k \mathcal{R}_k\right]\;.\label{FRGE}
\end{equation}
Here  Tr comprises a functional trace that includes an integration over spacetime, and $\Gamma_k^{(2)}$ represents the Hessian of the EAA with respect to the fluctuations. Importantly, this equation is fully non-perturbative, as no approximation or expansion has been performed in the derivation. Also, it is UV- and IR-finite wrt.  the eigenvalues of the d'Alembertian, due to the regulator $\mathcal{R}_k$ in the numerator and denominator, respectively.

This equation can be used to study the RG flow of the system in consideration. The first step consists in computing the beta functions $\beta_i$. One begins by specifying the space of functionals, called theory space, on which the FRGE  acts. Any element in this space can be written as a linear combination of basis functionals $\{P_i[\phi]\}$, with scale-dependent generalized couplings $U^i(k)$
\begin{equation}\label{basis}
\Gamma_k[\phi] = \sum_{i = 1}^\infty U^i(k) P_i[\phi]\;.
\end{equation}
 Geometrically, the FRGE defines a vector field on theory space $\mathbf{\beta}$, whose integral curves correspond to RG trajectories $k \to \Gamma_k$ parameterized by the RG scale $k$. This setup allows one to compute how the couplings $U^i(k)$ change with the scale $k$. The RG flow \eqref{FRGE} in the basis dependent component form basis \eqref{basis} reads:
 \begin{equation}
 	k	\partial_k	\sum_{i=1}^{\infty}  U^i (k)P_i[\phi]=  \frac{1}{2}\;\text{Tr}\;\left[ \left(\sum_{i=1}^{\infty}  U^i (k)P^{(2)}_i[\phi]+\mathcal{R}_k\right)^{-1}\;k\partial_k \mathcal{R}_k\right]\;.
 \end{equation}
 The RHS of this equation is an extremely complicated functional of $\phi$. However, the theory space, by its very definition,   contains all functionals that could possibly be produced by those algorithms, it follows that the $\phi$-dependence of $\frac{1}{2} \text{Tr}[\cdots]$ can be expanded in the basis $\{P_i [\phi]\}$:
 \begin{equation}
 	\frac{1}{2} \text{Tr}[\cdots]= \sum_{i=1}^{\infty}  b^i(U^1 , U^2, \cdots ;k)P_i[\phi]\;.
 \end{equation}
 Here the \textit{beta functions} $b^i$ arise expanding the trace on the RHS of the \eqref{FRGE} in terms of $\{P_i[\phi]\}$ and comparing the coefficients multiplying the same operator $P_i[\phi]$. Finally, it is customary to work with dimensionless variables in order to avoid rescaling of physical lengths. If the coupling $U^i$ has canonical dimension $d_i$ we  define the corresponding dimensionless coupling $ u^i (k) = U^i (k)k^{-d_i}$. Then, the $\beta$-functions describing the RG running of the dimensionless couplings are
 \begin{equation}
 	k\partial_k u^i = \beta^i(u^1, u^2, \cdots;k)\;.\label{beta1}
 \end{equation}

As a second step, the fixed point(s) can be determined, i.e. those values of the coupling constants $u^j_*$ for which all the beta functions vanish simultaneously 
\begin{equation}
\beta_i(u^j_*)=0 \; \qquad \forall i\;.
\end{equation}

Furthermore, as a third step, the critical exponents (or scaling exponents) $\theta_i$ at the fixed point(s) can be calculated by constructing the stability matrix $M$
\begin{equation}
M_i^j=\frac{\partial \beta_i}{\partial u^j}\Bigg|_{u^j=u^j_*}\; \qquad \to \qquad \theta_I=-\text{Eigen}\left(M_i^j\right)
\end{equation}
and computing its eigenvalues at the fixed point. Linearizing around a fixed point, these represent how the coupling constant scale in the vicinity of a fixed point
\begin{equation}
\lambda^j(k) = \lambda_*^j + \sum_I c_I v_I\left(\frac{k_0}{k}\right)^{\theta_I}\;.
\end{equation}
Here $k_0$ is a fixed scale, the $c_I$ are the constants of integration that can be expressed in terms of the initial conditions, and the $v_I$ are the right eigenvectors.
The critical exponents $\theta_I$ can be  related to the scaling of correlation functions. Consider for instance the two point function of a scalar field $\phi$ for a shift- and rotational-symmetric system. This will have the general form
\begin{equation}
G(x,y)=\langle \hat\phi(x)\hat \phi(y)\rangle \sim \frac{1}{|x-y|^\alpha}\;.
\end{equation}
The exponent $\alpha$ can be directly related to the scaling exponents $\theta_I$  by determining the scaling of the coupling constants. In the case of the two-point function this is related to the anomalous dimension $\alpha = d-2+\eta$ via the wave function renormalization $Z_k$.  For instance, close to  a fixed point
\begin{equation}
G_k(p) \sim \frac{1}{Z_k p^2} \overset{\text{at fixed point}}{\sim} \frac{1}{p^{2-\eta}}\;.
\end{equation}
These exponents can be  directly compared with lattice simulations or experiments \cite{Ambjorn:2024qoe}. Importantly, in the setup we aim to establish, such correlation functions and their exponents are constructed for gauge-invariant operators.

At this stage, our perspective can slightly shift from the conventional QFT approach: equipped with the renormalization group equation, we can adopt it as a constructive tool to define the QFT itself \cite{Manrique:2008zw,Manrique:2009tj, Reuter:2019byg}, specifically, by analyzing the behavior of  $\Gamma_k$ along the flow. 
In particular, if the flow associated with a given $\Gamma_k$ exhibits a regular behavior in both the $k \to \infty$ and the $k \to 0$ regimes, i.e., it possesses a well-defined $k \to \infty$  limit realized via a fixed point, then we can integrate down the flow and compute the $n$-th correlation functions from the full effective action by evaluating the $n$-th derivative of $\Gamma_{k \to 0}$ in the limit $k \to 0$.

The particular functional form of $\Gamma_k$ can be chosen depending on the field content under investigation and by the required symmetries. Ideally, one should include an infinite tower of those terms in order to study the entire theory space as in \eqref{basis}; however, this is often practically not feasible, because it would turn the flow equation \eqref{FRGE} into an infinite number of differential equations. Hence, a viable option is to systematically  truncate the operators entering in the ansatz of $\Gamma_k$, such as in a derivative expansion or in a vertex expansion \cite{Dupuis:2020fhh}, and analyze the stability of the result by refining the truncation. On the other hand, in certain cases, it is possible to perform infinite truncations at the functional level, as in $f(R)$ gravity \cite{Machado:2007ea,Falls:2016msz,Morris:2022btf}, where the EAA  is written as a general function of the Ricci scalar, capturing an infinite set of couplings within a closed functional form.

Recalling the ambiguity about the measure in Section \ref{sec:4}, we can argue in light of this new construction of the path integral, that it is the combination measure plus action which matters, as they always come together $\int \mathcal{D}\hat \phi e^{-S[\hat \phi]}$. Hence, in this new spirit, testing the properties of a given ansatz for $\Gamma_k$ is effectively equivalent to exploring different combinations of action and associated measure \cite{Manrique:2009tj,Bonanno:2025xdg,Held:2025vkd}.

As a final remark, in an attempt to establish the full connection with the bare theory $S$, the reconstruction problem was encountered \cite{ Manrique:2008zw,Manrique:2009tj,Morris:2015oca,Reuter:2019byg,Fraaije:2022uhg}. This refers to the challenge of reconstructing the bare action from a particular EAA, particularly when a UV fixed point governs the high-energy behavior as in $\Gamma_{k \to \infty} = \Gamma_*$. Unlike in perturbative theories in the FRG spirit, we don’t know the microscopic bare action $S_{\Lambda_{UV}}$ at the UV cutoff $\Lambda_{UV}$.  The reconstruction  problem relates to the fundamental question, whether  there has to be  a fundamental path-integral formulation behind  an effective (non-perturbative) theory and is crucial to be addressed in an attempt to connect with different fundamental theories of quantum gravity, such as with CQG.
In first one-loop investigations \cite{Manrique:2008zw,Manrique:2009tj} it was found that given a choice of UV regularization scheme and functional measure, any solution to the flow equation for the Effective Average Action, formulated without a fixed UV cutoff, can be associated with a corresponding regularized path integral. This path integral remains well-defined and exhibits a controlled limit as $\Lambda_{UV} \to \infty$, thereby providing a consistent non-perturbative definition of the theory.

\subsection{RG flows of gravity}
As mentioned in the introduction and in Section \ref{sec:2}, one of the strengths of this approach is that it can be implemented using the background field method without ever requiring an explicit specification of the background structure. After having introduced the flow equation, we can clarify how background independence is realized.

In eq. \eqref{FRGE} on the RHS the supertrace  has to be performed, over an operator consisting of the Hessian and the regulator kernel. By means of the background field method, in the context of dynamical gravity one chooses the kernel of the cutoff to regularize modes, typically over the background structure, i.e., 
\begin{equation}\label{eq:19}
\Delta S_k = \int \text{d}^d x \sqrt{\bar g} h_{\mu \nu} \mathcal{R}_k^{\mu \nu \rho \sigma}(\overline{\square})h_{\rho \sigma}\;,
\end{equation}
where $h$ is the fluctuation as specified in equation \eqref{eq:1}.
Evaluating the trace in the flow equation then boils down to compute the trace over the index structure and over the eigenvalues of the d'Alembert operator $\overline{\square} \equiv \bar g_{\mu \nu}\bar \nabla^\mu \bar \nabla^\nu$.

Crucially, such a trace can be performed without needing to specify on which background structure it is constructed on. The main tool to achieve this is the \textit{heat kernel expansion} \cite{DeWitt:1960fc,Christensen:1976vb,Barvinsky:1985an,Moretti:1999ez,Decanini:2005gt,DeWitt:2003pm,Parker:2009uva,Ferrero:2023xsf}. This is a very versatile framework to compute  the $\text{Tr}\; [e^{-s \overline{\square}}]$ or generalizations to non-minimal operators $\text{Tr}\;[\mathcal{O}\;e^{-s \overline{\square}}]$  \cite{Benedetti:2010nr,Groh:2011dw,Barvinsky:2021ijq,Ferrero:2023xsf}on a general $\bar g_{\mu \nu}$, where $s$ is  the heat kernel or proper time.

The first step to evaluate the trace consists in using the Schwinger proper time parametrization. This allows to  translate the trace of the inverse in the flow equation into the trace over the heat kernel:
\begin{equation}\label{reg}
\frac{	1}{\overline{\square}+ \mathcal{R}_k} = \int_0^\infty \text{d}s \; e^{-s( \overline{\square}+ \mathcal{R}_k)}\;.
\end{equation}
The heat kernel expansion can be then expressed for a general manifold as an expansion in power of the heat kernel time and curvature invariants encoded in the coefficients $a_n$
\begin{equation}
 e^{-s\overline{\Box}}= \left(\frac{1}{4\pi s}\right)^{d/2} \sum_{n = 0}^\infty s^n  \;a_n (x)\;.
	\label{earlytimes}
\end{equation}
The first three  coefficients in eq.\eqref{earlytimes} are
\begin{eqnarray}
	a_0 (x)& =& \mathbbm{1}\;,\\
	a_1 (x) &=& E+\frac{1}{6}\bar R\;  \mathbbm{1}\;,\\
	a_2(x) &=& \frac{1}{180}\left(\bar R_{\mu \nu \rho \sigma}\bar R^{\mu \nu \rho \sigma} +\bar R_{\mu \nu}\bar R^{\mu \nu} + \bar \nabla^2\; \bar R\right)\; \mathbbm{1} \nonumber\\
	&& + \frac{1}{2}\left( E+\frac{1}{6}\bar R\; \mathbbm{1}\right) ^2 + \frac{1}{12}[\bar \nabla_\mu, \nabla_\nu][\bar \nabla^\mu, \bar \nabla^\nu] + \frac{1}{6}\bar  \nabla^2\; \left(E+\frac{1}{6}\bar R\;  \mathbbm{1}\right)\;.
\end{eqnarray}
Here $E$ is a general endomorphism which can appear as a part of the Hessian.
These coefficients are found by solving the so called heat equation in terms of the Synge's world function \cite{Synge:1931zz} and the Van Vleck's determinant \cite{VanVleck:1928zz} and by solving order by order in the small $s$ expansion the heat equation. This practically translates  in recursion relations for the coefficients  simplified by the properties of the coincidence limit of the world function and its derivatives \cite{Ferrero:2023xsf}. 

Solving the flow equation requires tracing \eqref{earlytimes} and this will give
\begin{equation}
	\text{Tr} \left[ e^{-s\overline{\Box}}\right]=\left(\frac{1}{4\pi s}\right)^{d/2} \sum_{n = 0}^\infty s^n \int \text{d}^d x \sqrt{g} \;\text{tr} \;a_n (x) \;.
	\label{earlytrace}
\end{equation}
To first order in curvature this gives
\begin{equation}\label{smallR}
	\text{Tr} \left[ e^{-s\overline{\Box}}\right]=\left(\frac{1}{4\pi s}\right)^{d/2} \text{tr}({\mathbbm{1}})\int \text{d}^dx \sqrt{g}\;\left(1+ \frac{\bar R}{6}s + \cdots\right) \;,
\end{equation}
where $\text{tr} ({\mathbbm{1}})$ is the trace over the field space components we are computing the d'Alembertian of and the dots stand for different combinations of  higher powers in curvature monomials.
Importantly,  such an expansion perfectly fits with the approach of constructing the theory by starting with an ansatz on $\Gamma_k$, since it provides all the curvature invariants which can appear in such an ansatz.
Finally, comparing the operators appearing on the LHS and on the RHS of the flow equation \eqref{FRGE}, we can read off the beta functions for the coupling constants which were present in the original ansatz for $\Gamma_k$, without never specifying which particular value the operators take. In some sense, then, it is as we would had introduced a background but at the same time we are performing the computation on all possible backgrounds.

We emphasize here that the trace \eqref{smallR} is indeed IR divergent for $d=4$ and needs to be regularized. One common choice of the regulator is  the $\mathcal{R}_k$ introduced in \eqref{eq:19}, with a kernel constructed as a function of the background d'Alambertian. Alternatively, different kind of regulators for related flow equations have been considered in the literature, such as the Polchinski flow equipped with an UV regulator \cite{Polchinski:1983gv}  or the proper time flow equation \cite{Bonanno:2004sy,deAlwis:2017ysy}, equipped with a regulator directly in proper time. A regulator defined directly in proper time will be the method used also in this work in Section \ref{sec:6}, since it both overcomes some mathematical subtleties related to the Laplace  transform (Fourier transform in Lorentzian signature) in \eqref{reg} used to pass to the Schwinger proper time parametrization on one side, and it also bypasses the interpretational issues of integrating out modes in Lorentzian signature on the other side.

\bigskip

The effectiveness non-perturbative renormalization in gravity was directly tested in 1996 by Martin Reuter \cite{Reuter:1996cp}, who adopted the simplest ansatz for the EAA: the Einstein--Hilbert action, supplemented by a standard gauge-fixing term and the corresponding ghost term
\begin{equation}
\Gamma_k = \frac{1}{16 \pi G_N(k)} \int \text{d}^4x \sqrt{g} \left(2\Lambda(k)-R[g]\right)\; + \text{ gauge fixing }+ \text{ ghosts }\;.
\end{equation}
Here  $G_N(k)$ is the  scale-dependent Newton constant and $\Lambda(k)$ the scale-dependent cosmological constant. The gauge fixing chosen is typically a 
background covariant harmonic gauge
\begin{eqnarray}
	F^\mu
	&=& \frac{1}{\sqrt{16 \pi G_N(k)}}  \left( \bar{g}^{\mu \lambda} \bar{g}^{ \nu \rho}  - \frac{1}{2} \bar{g}^{\nu\mu}   \bar{g}^{\rho \lambda}     \right)  \bar{\nabla}_\nu  g_{\lambda \rho}\, ,\label{eq:gauge}
\end{eqnarray}
leading to the ghost operator ($c$ and $\bar c$ are the ghost and the antighost, respectively)
\begin{equation}
	(\mathcal{Q}_{\text{ FP}})^{\mu}\,_{\nu} c^{\nu} \equiv - \sqrt{16 \pi G_N(k) }\mathcal{L}_{c} F^{\mu} =  - \left(  \bar{g}^{\mu \lambda} \bar{g}^{ \nu \rho}  - \frac{1}{2}\bar{g}^{\nu\mu}   \bar{g}^{\rho \lambda}     \right)  \bar{\nabla}_\nu ( g_{ \rho \sigma} \nabla_{\lambda} c^{\sigma}  + g_{ \lambda \sigma} \nabla_{\rho} c^{\sigma}) \,,
\end{equation}
	which enters the gauge-fixing action
\begin{equation}
	S_{{\rm gf} }= \frac{1}{2 } \int {\rm d}^d x \sqrt{\bar{g}} F^{\nu}  \bar{g}_{\mu\nu} F^{\mu} \,,
\end{equation}
and the ghost action
\begin{equation}
	S_{\rm{gh}}= \int {\rm d}^d x\sqrt{\bar g}   \bar{c}_\mu (\mathcal{Q}_{\text{ FP}})^{\mu}\,_{\nu} c^{\nu}  \,,
\end{equation}
respectively.

\begin{figure}[h]
 	\centering
 	\includegraphics[width=.77\textwidth]{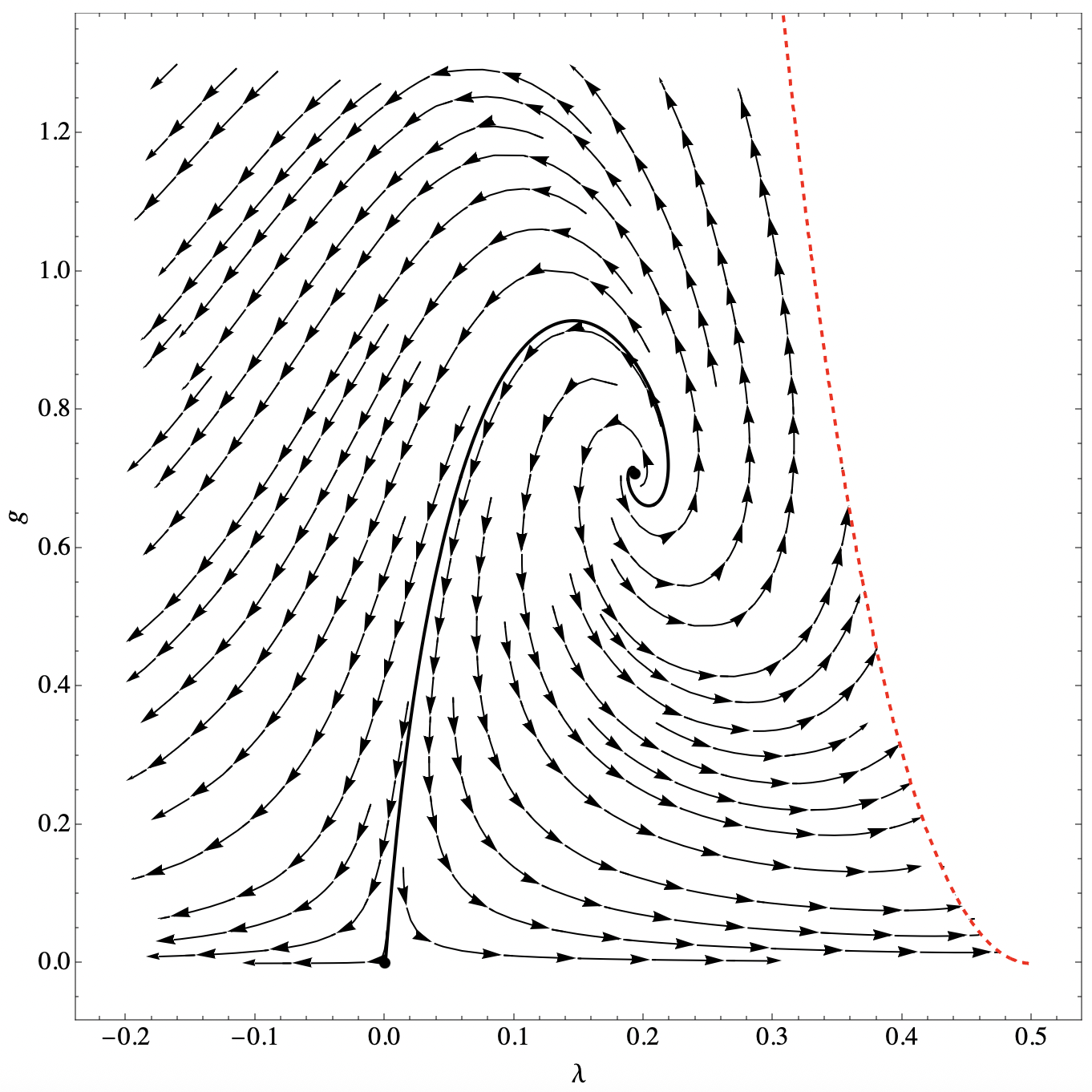}
 	\caption{Flow diagram in the $\lambda$-$g$ plane, dimensionless  $\Lambda(k)$ and $G_{N}(k)$. The IR fixed point at $\lambda_*, g_* =(0,0)$ and the UV fixed point $\lambda_*, g_* =(0.19, 0.71)$ are depicted with a black dot. All the trajectories  stem from the UV fixed point (the arrows' direction points towards decreasing RG scale $k$. The thick black line represent the trajectory which connects the UV to the IR fixed point. Furthermore, the flow cannot be trusted in regions beyond the red dashed line, since it becomes singular.}
 	\label{fig:1}
 \end{figure}

 The first analysis performed via the FRG revealed that the theory possesses a non-trivial UV fixed point and is asymptotically safe \cite{Souma:1999at}. In Figure \ref{fig:1} we report a flow diagram for the dimensionless $\lambda = \Lambda(k)/k^2$ and $g = G_N(k)k^2$ exhibiting an IR and an UV fixed point. The associated critical exponents are \begin{equation}
\theta_{1,2} = 1.47531 \pm 3.04322 \;i\;.
 \end{equation}
 Having a positive $\text{Re}\;\theta_{1,2}>0$, the two operators in the Einstein--Hilbert truncation are associated to two relevant directions, i.e., related perturbations  grow under the RG flow toward the IR. Furthermore, the critical exponents being complex conjugate is a manifestation of the spiraling of the trajectories around the UV fixed point \cite{Lauscher:2001ya}.
 For more details we refer the reader to Refs. \citen{Reuter:2001ag,Reuter:2012id}.
 
 \bigskip

At this stage let us mention that there are several excellent reviews and books on asymptotic safety, including Refs. \citen{Percacci:2017fkn,Reuter:2019byg,Bonanno:2020bil}, where more details and applications are reported. The program has found more evidence for such a fixed point, by refining the ansatz for  $\Gamma_k$ \cite{Falls:2014tra,Baldazzi:2023pep}, including field redefinitions \cite{Ihssen:2025cff} within the essential RG \cite{Baldazzi:2021orb}, by coupling different matter fields to gravity \cite{Dona:2013qba,Eichhorn:2018yfc}, by using different carrier fields than just the metric fluctuation \cite{Daum:2010zz,Daum:2013fu} and by studying bi-metric flows \cite{Becker:2014qya}. Important to mention is that asymptotic safety can also be tested via alternative RG methods \cite{Bonanno:2004sy, Falls:2024noj} or in a discrete setting by lattice gravity, as it is done in dynamical triangulations \cite{Loll:2019rdj,Ambjorn:2022naa} (for connections between discrete and continuous realizations we refer to Ref. \citen{Daum:2008gr,Reuter:2011ah,Ambjorn:2024qoe,Ambjorn:2024bud}) . As a side development aside from non-perturbative methods, evidence has been furnished, about the realization of the non-trivial RG UV fixed point within perturbation theory, by using background independent techniques, such as the proper time flow or dimensional regularization in a non-minimal subtraction scheme \cite{Kluth:2024lar,Falls:2024noj}.

\bigskip

Instead of going into those details, let us now deal with the topic from a different perspective, namely from a standpoint more familiar to the canonical approach. 
From a canonical standpoint, several concerns arise from this analysis
\begin{enumerate}
\item Typically, computations using FRG techniques are performed mainly in the Euclidean signature. 
\item In constructing the path integral, the issue of quantization in the UV regime has not properly been addressed, meaning that the corresponding quantum states have not been explicitly constructed.
\item It seems that the existence of asymptotically scattering states corresponding to free particles was assumed without carefully specifying boundary conditions for them. 
\item The integration measure of the path integral remains undefined. 
\item Crucially, the physical, observable degrees of freedom were not  distinguished from the unphysical, gauge-dependent ones.
\end{enumerate}    
These omissions raise important questions about the consistency and completeness of the quantization procedure.

Let us comment on the first four points more generally and focus more on the last point in the next section. 

The issue of the theory's signature has long been a cause for concern. It is only in the last couple of years that flourishing attempts  to perform a Lorentzian computation have been pushed forward and are now well-established \cite{Manrique:2011jc,Biemans:2016rvp,Bonanno:2017gji,Fehre:2021eob, Banerjee:2022xvi,DAngelo:2022vsh,DAngelo:2023wje,DAngelo:2025yoy,Saueressig:2025ypi,Ferrero:2024rvi}. The issue of going Lorentzian is twofold: on one side there is the technical challenge to compute the trace as described above for the Lorentzian flow equation; on the other side, several interpretational issues  regarding the meaning of the cutoff arise due to the ambiguity of regularizing spacelike and timelike momenta (no well-defined integrating-out ordering is given). As for the first challenge, among the most promising methods, it has been  proposed to exploit methods well established in the Algebraic QFT framework \cite{DAngelo:2022vsh} on one side, and to extend the heat kernel (or better said Schrödinger) traces to Lorentzian signature on the other side (see Refs. \citen{Estrada:1997qt,Moretti:1999ez,Strohmaier:2023wsv} for a connection between the two methods and \cite{Banerjee:2024tap,Banerjee:2025jhm} for a discussion on the Schrödinger group from an analytically continued lapse). In fact, recently, a way how to solve the resulting flow equation by means of a cutoff in heat kernel space has been pushed forward \cite{Thiemann:2024vjx, Ferrero:2024rvi, Ferrero:2025idz}.  The heat kernel expansion in this context reads:
\begin{equation}\label{HKLorentz}
	\text{Tr} \; \left(e^{is \overline{\square}} \right)=\frac{\text{tr} (\mathbbm{1})e^{i \frac{\pi}{4}\text{sgn}(s)(2-d)}}{(4 \pi |s|)^{d/2}}\int \text{d}^dx\sqrt{-g}\;\left(1+ \frac{i}{6}\bar R s + \cdots\right)
\end{equation}
Hence, we notice that the price to pay for such a Lorentzian heat kernel expansion is that the coupling constants become generically complex. For a discussion about this new approach we refer the reader to Refs. \citen{Thiemann:2024vjx, Ferrero:2024rvi}.
What the second Lorentzian issue is concerned, such an ambiguity is unavoidable but it also sheds  light on some interpretational aspects which in Euclidean signature were hidden. For instance, it seems natural to think that the selection of which modes have to be integrated out is dependent on the experimental setting. This would introduce an observer dependence into the RG process, invalidating any possibility of having a unique Lorentzian flow. This was investigated in the context of de Sitter spacetime by analyzing the running spectral flow \cite{Ferrero:2022hor} and opens the avenue for more interesting investigations on physical, relational aspects of the RG methodology in Lorentzian signature.

The EAA approach adopts a constructive formulation of the path integral. In essence, much like in lattice field theory, one begins by proposing an ansatz for a scale-dependent, ``running" action. The path integral is then evaluated by integrating over the most straightforward choice of variables, namely the fields themselves.
A critical requirement in this framework is that the path integral
$
\int \mathcal{D}g_{\mu \nu}e^{-S[g]},
$
comprising both the action and the integration measure, respects the fundamental symmetries of the theory. In the context of quantum gravity, this specifically means preserving diffeomorphism invariance \cite{Bonanno:2025xdg}. Ensuring this symmetry is maintained at the quantum level is essential for the consistency and physical viability of the theory, as it reflects the underlying coordinate independence of GR. In this context, the challenges typically associated with quantization, such as the explicit construction of quantum states, the imposition of boundary conditions, and the precise definition of the integration measure, can be effectively partially sidestepped. This is possible because the approach focused until now on the flow of the EAA in a constructive way, rather than starting from a full canonical quantization \cite{Manrique:2008zw, Manrique:2009tj}.

However, when it comes to defining candidate observables, especially in a Lorentzian spacetime, the above mentioned  issues need to be addressed. At this stage, the connection to the Hamiltonian (canonical) formalism becomes particularly compelling and explanatory and  will be the topic of Section \ref{sec:6}.

\section{Canonical Quantum Gravity:  from relational observables to background-independent quantization}\label{sec:5}
The starting point of CQG is the Hamiltonian formulation of Einstein’s equations, typically expressed in terms of the ADM (Arnowitt-Deser-Misner) decomposition \cite{Arnowitt:1962hi} which foliates spacetime into a family of spatial hypersurfaces. The canonical variables consist of the $(d-1)$-metric $q_{ij}$ on a spatial slice and its conjugate momentum  related to the extrinsic curvature of the slice.

As shown in Section \ref{sec:3}, the phase space of the theory is subject to a set of first-class constraints: the scalar (or Hamiltonian) constraint and the vector (or diffeomorphism) constraints. These reflect the underlying spacetime diffeomorphism invariance and must be imposed as operator conditions on the quantum states. 

In the next subsections, we will outline the main procedure for handling a Hamiltonian constrained system and its quantization and we will derive the path integral  of CQG.

\subsection{Reduced Hamiltonian and its quantization}
As showed in Section \ref{sec:3}, in order to define the time evolution, we need to review elements of canonical quantization of constrained systems \cite{Henneaux:1994lbw}.

GR is a constrained system with a set of first-class constraints $C_j$
\begin{equation}
\{C_j,C_k\}=f_{jk}\;^l\;\; C_l\;,
\end{equation}
where the $f$ are the structure functions that are not constant on phase space in the case of GR. Let us consider a set of degrees of freedom constituted by  $(x^i, q^a)$ and their respective momenta $(y_i, p_a)$.  The reduced phase space can be constructed in two conceptually equivalent approaches \cite{Thiemann:2004wk}: one involves identifying relational Dirac observables that are invariant under gauge transformations \cite{Dirac:1958sc}, while the other proceeds by imposing explicit gauge-fixing conditions to eliminate redundant degrees of freedom $(y_i,x^i)$ and keep only the physical degrees of freedom $(p_a,q^a)$. We will detail in this section the second procedure. In both cases first we solve the constraints $C_j=0$ 
for the momenta $y_j$ to arrive at equivalent sets of constraints
\begin{equation}
\hat{C}_j=y_j+h_j(p,q;\;x)\;.
\end{equation}
Note that the dependence of the functions $h_j$ on $x$ can be in general quite complicated. To deal with that, one can  select a split of the canonical coordinates into the two sets in such a way to avoid these complications as much as possible.
Importantly, the virtue of the equivalent set of constraints $\hat C_j$ is that they should be Abelian, i.e., they have vanishing Poisson brackets among each other \cite{Dittrich:2005kc}. Hence, there are no  operator-ordering ambiguities and we can proceed with the  construction of the so-called true or physical (or reduced) Hamiltonian.

In the gauge-fixing approach we consider the reduced phase space (which comprehends only physical degrees of freedom) as  coordinatized by the so called ``true degrees of freedom''
$(P,Q)$ (capital variables are coordinate functions on the phase space $P^a(z)=p^a$).  The ``gauge degrees of freedom'' $(Y,X)$ are discarded from the description from scratch.

Consider then the coordinate functions  $k^j:\; \mathbb{R}\to \mathbb{R};\;\; t\mapsto c^j(t)$ which are constants 
on the phase space and a one-parameter set of gauge-fixing conditions 
\begin{equation}\label{eq:25}
G^j(t):=X^j-k^j(t)   \;.
\end{equation}
Next consider a gauge transformation generated by the constrained 
Hamiltonian $\hat{C}(s)=s^j \hat{C}_j$, where $s^j$ are Lagrange multipliers. There 
is a residual one-parameter  family of gauge transformations allowed   
that stabilize the gauge condition $G^j(t)=0$, i.e. which satisfies $\frac{d}{dt} G^j (t) = 0$, namely
\begin{equation}
\frac{d}{dt} G^j(t)=\partial_t G_j(t)+\{\hat{C}(s),G^j(t)\}
=-\dot{k}^j(t)+s^j=0\;,
\end{equation}
which fixes the Lagrange multiplier $s^j$. We denote the fixed values 
\begin{equation}
s^j_\ast=\dot{k}^j(t)\;,\;
X^j_\ast=k^j(t)\;,\;
Y_j^\ast=-h_j(Q,P;\;k(t))\;.
\end{equation}
Now, let $F$ be a function of the true degrees of freedom $q,p$ only. Then 
the reduced or physical Hamiltonian $H_\text{red}(t)=H_\text{red}(q,p;t)$ must satisfy
\begin{equation}
\{H_\text{red}(t),F\}(q,p):=(\{\hat{C}(s),F\}_{s=s_\ast,x=x_\ast,y=y^\ast})(q,p)
=\dot{k}^j(t)\{h_j(.,.;c(t)),F\}(q,p)\;,
\end{equation}
whence 
\begin{equation}
H_\text{red}(q,p;t)=\dot{k}^j(t)\; h_j(q,p;k(t))\;.
\end{equation}
It is important to emphasize at this stage that the physical meaning of the true degrees of freedom and the reduced Hamiltonian depends on the chosen gauge fixing via \eqref{eq:25}.
The time evolution of the physical degrees of freedom is then governed by the reduced Hamiltonian associated with the gauge condition $G^j(t) = 0$. We also highlight again that this procedure is isomorphic to the construction of Dirac relational observables.

\bigskip

Given the reduced Hamiltonian, we can proceed with the quantization.
Canonical quantization requires constructing a representation of the canonical commutation relations along with the corresponding *-algebra structure\footnote{A *-algebra is an algebraic structure equipped with an operation called an involution, denoted by the star, that generalizes complex conjugation or the adjoint operation on operators.}
\begin{equation}
	[P_a,Q^b] = i\delta^b_a \mathbbm{1}\;.
\end{equation}
Since  the $P$ and $Q$ operators are unbounded, it is convenient to pass to the  Weyl elements
\begin{equation}
	W[j,g] =: e^{ij_a Q^a+ i g^aP_a}\;,
\end{equation}
where $j_a$ and $g^a$ are smearing functions.
The Weyl elements generate the associated Weyl algebra $\mathcal{A}$, which forms a *-algebra. We study cyclic representations $\Omega$ of the algebra $\mathcal{A}$, each built from a special state $\omega$, exploiting the fact there is a one-to-one correspondence between such states and representations.

The state $\omega$ leads to a set of data called GNS data (Gelfand–Naimark–Segal), consisting of a Hilbert space $\mathcal{H}$, a unit vector $\Omega \in \mathcal{H}$ and a *-representation $\rho$ of $\mathcal{A}$ as operator on $\mathcal{H}$.
The state  $\omega$ is connected to the GNS data by the formula:
\begin{equation}
	\omega (a) = \langle \Omega, \rho (a) \Omega\rangle_{\mathcal{H}}, \qquad \forall a \in \mathcal{A}\;.
\end{equation}
This means that $\omega$  then looks like an  expectation value and furnishes the correlation functions for the operators $\rho$ when applied to the vector $\Omega$,  like in standard QFT or in Algebraic QFT (AQFT).

Furthermore, constructing a quantum version of the Hamiltonian $H_\text{red}$ presents several challenges. These include: ambiguities in operator ordering when translating classical expressions into quantum operators, UV divergences arising from the need to sum over infinitely many high-frequency modes, and IR divergences that appear when dealing with non-compact spatial manifolds.

To manage the last two difficulties, it is standard practice in constructive quantum field theory \cite{Glimm:1987ng} to introduce regularizations. This typically involves imposing a UV cutoff $\Lambda_\text{UV}$ by restricting the modes to be smaller than $\Lambda_\text{UV}$, effectively limiting the contribution of high-energy modes \cite{Thiemann:2020cuq,Lang:2017beo}. Simultaneously, an IR cutoff is applied by compactifying the spatial manifold, commonly by placing the system on a finite-size torus. One then constructs the GNS data on such a regularized system $(\mathcal{H}_{\Lambda_\text{UV}}, \Omega_{\Lambda_\text{UV}},\rho_{
	\Lambda_\text{UV}} )$.

\subsection{Path integral formulation}
We now turn to a path integral framework.\footnote{We refer the reader to Refs. \citen{Isham:1984sb,Isham:1984rz,Lee:1990nz,Guven:1991eq} for  historical discussions on the subtleties relating constrained system and path integral formulation for gravitational theories.} We start again from some  regularized state $\omega_{
	\Lambda_\text{UV}}$, some (physical) Hamiltonian $H_{
	\Lambda_\text{UV}}$ and the related GNS data.
 Suppose that $H_{
 	\Lambda_\text{UV}}$ is bounded from below  and has a unique ground state $\Omega_{
 0,	\Lambda_\text{UV}}$, i.e. $H_{
 	\Lambda_\text{UV}} \Omega_{0,
 	\Lambda_\text{UV}} = 0$.  We are interested in constructing the time ordered $n$-point correlation functions.
Starting from the regularized vacuum expectation value of time-ordered Weyl operators we define:
 \begin{equation}
 	C_{n,  \Lambda_{\text{UV}}}((t_1,j^1),..,(t_n,j^n)) :=
 	\braket{\Omega_{0,\Lambda_{\text{UV}}} | \mathcal{T}\left[W_{t_n}(j^n,0)\cdots W_{t_1}(j^1,0)\right] | \Omega_{0, \Lambda_{\text{UV}}}}_{\mathcal{H}_{\Lambda_{\text{UV}}}} \;,
 \end{equation}
 where $W_t(j,g):=U_{ 	\Lambda_\text{UV}}(t) W(j,g) U_{ 	\Lambda_\text{UV}}(t)^{-1}$ and $U_{ 	\Lambda_\text{UV}}=\exp(-it H_{ 	\Lambda_\text{UV}})$. The time-ordering symbol $\mathcal{T}$ arranges operators according to their time arguments, placing those with later times to the left and earlier ones to the right.
 Deriving by
 $j^k_{a_k},\; k=1,..,n;\; a\le  	\Lambda_\text{UV}$ and evaluating at $j^1=\cdots=j^n=0$ yields
 \begin{equation}
 	C_{n,\Lambda_\text{UV}}((t_1,a_1),..., (t_n,a_n)) := 
 	\braket{\Omega_{0,\Lambda_\text{UV}} | \mathcal{T}\left[Q^{a_n}(t_n)\cdots Q^{a_1}(t_1)\right] | \Omega_{0,\Lambda_\text{UV}}}_{\mathcal{H}_{\Lambda_{\text{UV}}}}\;.
 \end{equation}
 One can easily show that the generating functional is
 \begin{equation}\label{34}
 	\chi_{\Lambda_{\text{UV}}}(j) :=
 	\braket{\Omega_{0,\Lambda_{\text{UV}}} | \mathcal{T}\left[ e^{i\int \text{d}t\, j_a(t) Q^a(t)}\right] | \Omega_{0,\Lambda_{\text{UV}}}}_{\mathcal{H}_{\Lambda_{\text{UV}}}}\;.
 \end{equation}
 To regularize time evolution and to take into account the situation where no access to the vacuum state is given, we rewrite the vacuum expectation value as a limit
 \begin{equation}\label{eq.43}
 	\braket{\Omega_{0,\Lambda_{\text{UV}}}  | \;\cdot \;| \Omega_{0,\Lambda_{\text{UV}}} } = \lim_{T \to \infty} \frac{\braket{\Omega_{\Lambda_{\text{UV}}} | U(T) \; \cdot \; U(-T) | \Omega_{\Lambda_{\text{UV}}}}}{\braket{\Omega_{\Lambda_{\text{UV}}} | U(-2T) | \Omega_{\Lambda_{\text{UV}}}}}\;.
 \end{equation}
 Performing a  time-slicing (\( \epsilon_N = T/N \)) and inserting resolutions of identity, the numerator of \eqref{eq.43} becomes:
 \begin{equation}
 	\begin{aligned}
 		Z_{\Lambda_{\text{UV}},T,N}(j) =& \langle\Omega_{\Lambda_{\text{UV}}} | 
 		\underbrace{e^{i(T-t_{N-1}) H_{\Lambda_{\text{UV}}}}}_{\text{propagation}}  		\underbrace{e^{i\epsilon_N j_a(t_{N-1}) Q^a}}_{\text{source insertion}}e^{i\epsilon_N H_{\Lambda_{\text{UV}}}}  
 		\cdots \\
 		&\qquad\qquad\cdots e^{i\epsilon_N H_{\Lambda_{\text{UV}}}}
 		\underbrace{e^{i\epsilon_N j_a(t_{-N}) Q^a}}_{\text{source insertion}}
 		\underbrace{e^{i(t_{-N}-(-T)) H_{\Lambda_{\text{UV}}}} }_{\text{propagation}}
 		| \Omega_{\Lambda_{\text{UV}}}\rangle\;.
 	\end{aligned}
 \end{equation}
 
 We can now insert complete sets of states and write the path integral in phase space form. The generating functional can be rewritten symbolically as:
 \begin{equation}
 	\label{eq:Zfunctional}
 	Z(j) = \int \mathcal{D}p \, \mathcal{D}q 
 	\, \underbrace{\overline{\Omega(q(\infty))} \Omega(q(-\infty))}_{\text{initial/final data}}
 	\, \underbrace{e^{i\int \text{d}t\, j_a(t) q^a(t)}}_{\text{source term}}
 	\, \underbrace{e^{-i\int \text{d}t\, (p_a \dot{q}^a - H(q,p))}}_{\text{Hamiltonian phase}}
 \end{equation}
 
 \vspace{-2.0em}
 
 \begin{center}
 	\begin{tikzpicture}[remember picture, overlay]
 		\draw[decorate, decoration={brace, amplitude=8pt, mirror}, yshift=-4pt]
 		(-3,0) -- (-0.1,0) node [black,midway,yshift=-12pt] {\footnotesize GNS data};
 		
 		\draw[decorate, decoration={brace, amplitude=8pt, mirror}, yshift=-4pt]
 		(0,0) -- (2.1,0) node [black,midway,yshift=-12pt] {\footnotesize source};
 		
 		\draw[decorate, decoration={brace, amplitude=8pt, mirror}, yshift=-4pt]
 		(2.2,0) -- (5.2,0) node [black,midway,yshift=-12pt] {\footnotesize Lagrangian (dynamics)};
 	\end{tikzpicture}
 \end{center}
 \vspace{+2em}
 Note that in the last step we have taken the limits of vanishing  $N$, $T$ and  $\Lambda_\text{UV}$ cutoffs, i.e., we sent them to infinity, in the specified order.

 This completes the heuristic transition from the constrained system and the algebra of operators  to the path integral,  resulting in a schematic structure of the generating functional \( Z(j) \).	We emphasize at this point that the source term was denoted by $j$, in analogy to the source in QFT as in equation \eqref{eq:Z}, and that the Hamiltonian $H(q,p)$ will be identified with the reduced Hamiltonian $H_\text{red}$ constructed in the previous subsection.

This expression has indeed to be compared with \eqref{eq:Z} (without the regulator term).
It is important to emphasize that the path integral in equation \eqref{eq:Zfunctional} is  formulated over phase space, meaning both configuration variables and their conjugate momenta are included in the integration. This is distinct from more familiar path integrals in AS, that are defined purely over configuration space and where the measure has not to be specified (see Section \ref{sec:3}).

In some cases it is possible to integrate out the momenta to arrive at a configuration space path integral
\begin{equation}\label{eq:27}
	Z(j)=\int\; [dq]\; \mathcal{M}[q]\;\overline{\Omega(q(\infty))}\;\Omega(q(-\infty))
	\;
	e^{i\int\; dt\; j_a(t) q^a(t)} \; 
	e^{-\int\; dt L(q(t), \dot q (t))} \;,
\end{equation}
where $\mathcal{M}[q]$ represents the resulting non-trivial measure.
However, this can only be done when the Hamiltonian  depends on the momenta in a sufficiently simple way, typically at most quadratically \cite{Kunstatter:1991qe}. Even then, the integration may alter the structure of the path integral measure, so that the resulting configuration space measure is no longer a simple Lebesgue-type measure but is modified in a non-trivial way.
Additionally, it is important to observe that the partition function $Z(j)$ depends not only on the function $j$, but also on the choice of asymptotic cyclic vector $\Omega(\pm \infty)$, or equivalently, on the state $\omega$. \footnote{This is similar to what it is described in the AQFT formalism \cite{DAngelo:2022vsh}, while in the standard QFT formalism the asympotic state prescription implicitly amounts to choose asymptotically free states like in the construction of the $S$-matrix.} This dependence reflects the fact that in the operator formalism, the path integral is computing expectation values with respect to a specific quantum state, encoded through the GNS  construction.
\bigskip

Let us briefly summarize the situation. Starting from the fundamental theory formulated as a Hamiltonian system, we have derived a partition function by quantizing the theory in the spirit of the reduced phase space approach and we have identified the physical states. This partition function can serve now  as the basis for further analysis using FRG methods introduced in Section \ref{sec:4}. In particular, it allows us to investigate whether the resulting quantum theory exhibits asymptotic safety, i.e., whether it admits a non-trivial UV fixed point that ensures its predictive consistency at all  scales. Hence, the spirit here differs slightly from the approach discussed in the previous section regarding how the EAA is typically ansatzed, with the purpose to match IR observations, like in the case of GR by means of Einstein--Hilbert action. In this case, it is the underlying microscopic Hamiltonian theory that dictates the structure of the EAA. 

The advantage of this method lies in the fact that the path integral is constructed solely over physical degrees of freedom. As a result, applying an RG analysis directly corresponds to computing correlation functions and critical exponents of  physical observables. Importantly, within the reduced phase space framework, these observables are specific only to the given reference system, i.e., only to a given gauge fixing. In order to connect to different observables measured by different observers one should unfold the procedure and relate the different observers. This can however result in a challenging task, depending on the model in consideration.

Finally, we observe that, when starting from a Hamiltonian formulation, only a Lorentzian path integral is in general naturally well-defined, although in some particular cases the Wick rotated Euclidean version might be considered.
The derivation of the generating functional  leading to \eqref{eq:27} in Euclidean signature can be performed in parallel fashion, Wick rotating in time and effectively having the $i$-factors being absorbed.
In fact, in the following section, we will show under which circumstances it is possible to pass from the canonical, Lorentzian theory to an Euclidean path integral. We will demonstrate that, in certain specific cases, it is  possible to  perform an analytic continuation in a controlled way, even when the underlying Hamiltonian theory remains fundamentally Lorentzian. Whenever this is not the case, we will employ the novel method for performing Lorentzian FRG integrations via the Lorentzian heat kernel.

\section{Towards a cohesive approach for predictive Quantum Gravity }\label{sec:6}
Having derived a path integral formulation for the time evolution of physical observables given in \eqref{eq:27}, we are in the position to compute their running and carry out the renormalization procedure outlined in Section \ref{sec:4}.
Let us recall the differences from the standard QFT renormalization and from the AS approach: rather than using the conventional Faddeev--Popov procedure, which requires gauge-fixing and the introduction of ghosts term and which leads to gauge-dependent fixed points and critical exponents \cite{Benedetti:2010nr,Gies:2015tca, Bonanno:2025tfj, DAngelo:2025yoy}, we now work with a path integral in which the unphysical degrees of freedom have been eliminated from the outset. Moreover, since we have started from the physical Hamiltonian, we have direct access to the quantization procedure, the associated physical states, and the corresponding measure. The latter two appear explicitly in the path integral formulation and will play an important role in the flow of observables and their renormalization.

First of all, to concretely implement this program, we require a material reference system that allows us to eliminate the unphysical degrees of freedom from our formulation. Importantly, the introduction of this  reference system - aside from physical interpretational aspects \cite{Rovelli:1990pi, Giesel:2007wi, Giesel:2012rb, Thiemann:2023zjd} - alters the physical content of the theory, making it distinct from pure GR. However, the phase space is reduced in such a way that the material system serves solely to deparametrize the theory and does not itself appear among the physical observables.

\subsection{First examples of contact}
Recently, two different convenient examples of material reference systems have been used to exemplify this procedure:
\begin{enumerate}[label=(\alph*)]
\item $d$ scalar fields  coupled to gravity via a derivative interaction \cite{Ferrero:2024rvi}:
\begin{equation}
S_\text{S}=\frac{1}{16 \pi G_N}\; \int\; \text{d}^dx\; \sqrt{-g}\;\left(R[g]-2\Lambda-\frac{G_N}{2} 
S_{IJ}\; g^{\mu\nu}\;\phi^I_{,\mu}\;\phi^J_{,\nu}\right)\;.
\end{equation}
Here the Greek indices indicate the spacetime components while the capital Latin letters $I = 0,1,\cdots, d$). Furthermore $S^{IJ}$ is a general matrix of coupling constants. As a  minimal choice one could choose it proportional to a Kronecker  $\delta_{IJ}$, reducing the coupling constants to only one.
\item Gaussian dust matter fields \cite{Ferrero:2025idz}
\begin{equation}
	S_\text{GD}=\frac{1}{16 \pi G_N}\; \int\; \text{d}^dx\; \sqrt{-g}\;\left(R[g]-2\Lambda-\frac{\rho}{2}\;(g^{\mu\nu} \; 
	T_{,\mu}\; T_{,\nu}+1)+g^{\mu\nu}\;T_{,\mu}\; (W_j\; S^j_{,\nu})\right)\;.
\end{equation}
where  the minuscule Latin letters $j=1,..,d-1$. Thus, the action depends on $2d$ scalar fields $(T,S^j),\;(\rho,W_j)$. The variables are chosen to represent the physical coordinates: the $T$ scalar field serves as the time reference, it labels the proper time along the dust flow lines and effectively provides a material clock for the spacetime; the $S^j$ triplet of scalar fields labels  the spatial coordinates along the dust congruence. The variables $\rho$ and $W_j$ appear as Lagrange multipliers. In particular $\rho$ is the rest-mass density of the dust, enforcing the normalization of the dust four-velocity, while the $W_j$ serve to enforce the condition that the dust flow lines are non-intersecting and that the spatial coordinates are transported along the dust flow. More precisely, they ensure that the spatial labels 
remain constant along the dust worldlines.

The model is called dust since by means of the equations of motion we find a pressure-less perfect fluid. The equations of motion for $\rho$ and $W_j$  are respectively
\begin{equation}
g(U,U)=-1, \qquad g(U,V_j)=0\;,
\end{equation}
where we defined
\begin{equation}
U^\mu=g^{\mu\nu} T_{,\nu},\qquad V^\mu_j=g^{\mu\nu} S^j_{,\nu}\;.
\end{equation}
This  reveals that $U$ is a unit timelike geodesic tangent, i.e., $\nabla_U U=0$.
The equations of motion
$T,S^j$  yield the conservation 
equations
\begin{equation}
 \nabla_\mu(\rho U^\mu+W^j V_j^\mu)=0\;, \qquad\nabla_\mu (W_j \;U^\mu)=0\;.
\end{equation}
As a consequence the energy momentum tensor $T^{\mu\nu}=\rho U^\mu U^\nu+2\;W^j V_j^{(\mu}\; U^{\nu)}$ 
is conserved. Furthermore, the pressure
\begin{equation}
p=\frac{1}{d-1}(g^{\mu\nu}+U^\mu U^\nu)\; T_{\mu\nu}=0
\end{equation} vanished identically. 
\end{enumerate}

\bigskip
Let us confront now the two models considered. The two models represent different material frames; in both cases the physical variables are $(p_a, q^a) \equiv(p_{ab}, q^{ab})$, i.e., the spatial metric and its conjugate momentum. The unphysical ones instead are those related to the four scalar fields  $(y_i, x^i) \equiv (\pi_I, \phi^I)$ and to the eight  fields forming the dust $(y_i, x^i) \equiv (\{\pi_\rho, \pi_{W}^j, \pi_T, \pi _{S,j}\}, \{\rho, W_j, T, S^j\})$, respectively. Hence starting from the reduced phase space quantization, we will obtain different states and ultimately, due to the different deparametrization, a different generating functional. In particular, the dependence on the conjugate momentum will be different, leading to a different expression $\mathcal{M}$ in \eqref{eq:27}, once the momentum has been integrated out.

\bigskip

In the case of the four scalar fields (a) when $d=4$, when integrating out the conjugate momenta from \eqref{eq:Zfunctional}, the specific form of the physical Hamiltonian compels us to work in Lorentzian signature. The resulting path integral is
\begin{equation}
Z_\text{S}[j]=\int\; \mathcal{D}q\; \mathcal{M}[q]\; 
\overline{\Omega[Q(\infty)]}\; \Omega[Q(-\infty)]e^{-i S[g]\;}
\;e^{i\int\; \text{d}^dx \; j^{ab} q_{ab}},
\end{equation}
\begin{equation}\label{S1}
S[g]=
\frac{1}{16 \pi G_N}\int\;\text{d}^dx\sqrt{-g}\left(R[g]-2\Lambda-\frac{G_N}{2}
g^{\mu\nu}\;S_{IJ}\;\;k^I_{,\mu}\;k^J_{,\nu}\right)\;.
\end{equation}
Here the term containing $k^I$  is the “reduction term” reminiscent of the deparametrization or gauge fixing in Section \ref{sec:5} (see \eqref{eq:25}). Symbolically, it replaces the usual
gauge fixing term. In addition to this, we observe that the measure term could be also translated into a ghost integral using Faddeev--Popov's method. We recall the reader, that the action $S$ in \eqref{S1} has been read off from the Legendre transform of the reduced Hamiltonian obtained from reducing the phase space with respect of the four scalar fields \cite{Giesel:2012rb,Ferrero:2024rvi}.

Furthermore, by exploiting the ADM relations to connect $q_{ab}$ to $g_{\mu \nu}$ via the lapse function  $N$  and the shift function  $N^a$ in \eqref{gmunu}
one can change variables in  the functional integration in the path integral in order to obtain  an integration over $[dg]\mathcal{M}'[g]$, where the measure gets now accordingly modified. This comes with the advantage that we can use standard covariant heat kernel methods as in \eqref{HKLorentz} to compute the traces in the flow equation \eqref{FRGE}. 

It turns out to be particularly convenient to specialize to the gauge $\phi^I=k^I$ such that $k^I_{,\mu}=$ const.
and such that $S_{IJ} k^I,_{\mu} k^J,_{\nu}=
\kappa \delta_{\mu\nu}$ with $\kappa>0$ the only remaining coupling constant \cite{Ferrero:2020jts}. Considering this as an ansatz for the EAA, promoting the coupling constants to be RG scale-dependent and starting with the Einstein--Hilbert truncation in the canonical derivation, the EAA reads
\begin{equation}
\Gamma_k=
	\frac{1}{16 \pi G_{N}(k)}\int\;\text{d}^dx\sqrt{-g}\left(R[g]-2\Lambda(k)-\frac{G_N(k) \kappa(k)}{2}\;g^{\mu\nu}\; \delta_{\mu\nu}\right)\,.
\end{equation}
To simplify the initial analysis, the authors assumed constant physical states $\Omega$ and, for this first investigation, neglected the effects of the non-trivial measure $\mathcal{M}'$ that would otherwise enter the path integral. This approximation allowed for a more tractable computation while setting the stage for more refined treatments where the full structure of the measure will be properly incorporated.

In order to analyze the model, the Lorentzian heat kernel methods introduced in Section \ref{sec:5} were used. For this model, working in Lorentzian signature is compulsory: when the momenta get integrated out from the phase space path integral, the signature is dictated by the requirement of balancing the correct imaginary $i$ factors.

As a choice for the cutoff kernel in \ref{eq:19}, this was built directly in proper time as
\begin{equation}\label{cutoffpt}
\mathcal{R}_k^{\mu \nu \rho\sigma} (\overline{\Box})= K^{\mu \nu \rho \sigma} G_N(k)^{-1}k^2\; \int_0^\infty \text{d}s \;e ^{i \overline{ \Box} s}\;e^{-s^2-s^{-2}}\;,
\end{equation}
where $K^{\mu \nu \rho \sigma}$ represents a tensorial structure, typically a DeWitt metric.
This choice turned out to be particularly convenient, since it regularizes both $s \to \infty$ and $s\to 0$ \cite{Thiemann:2023yup,Neuser:2023pnr}.

\begin{figure}[h]
	\centering
	\includegraphics[width=.48\textwidth]{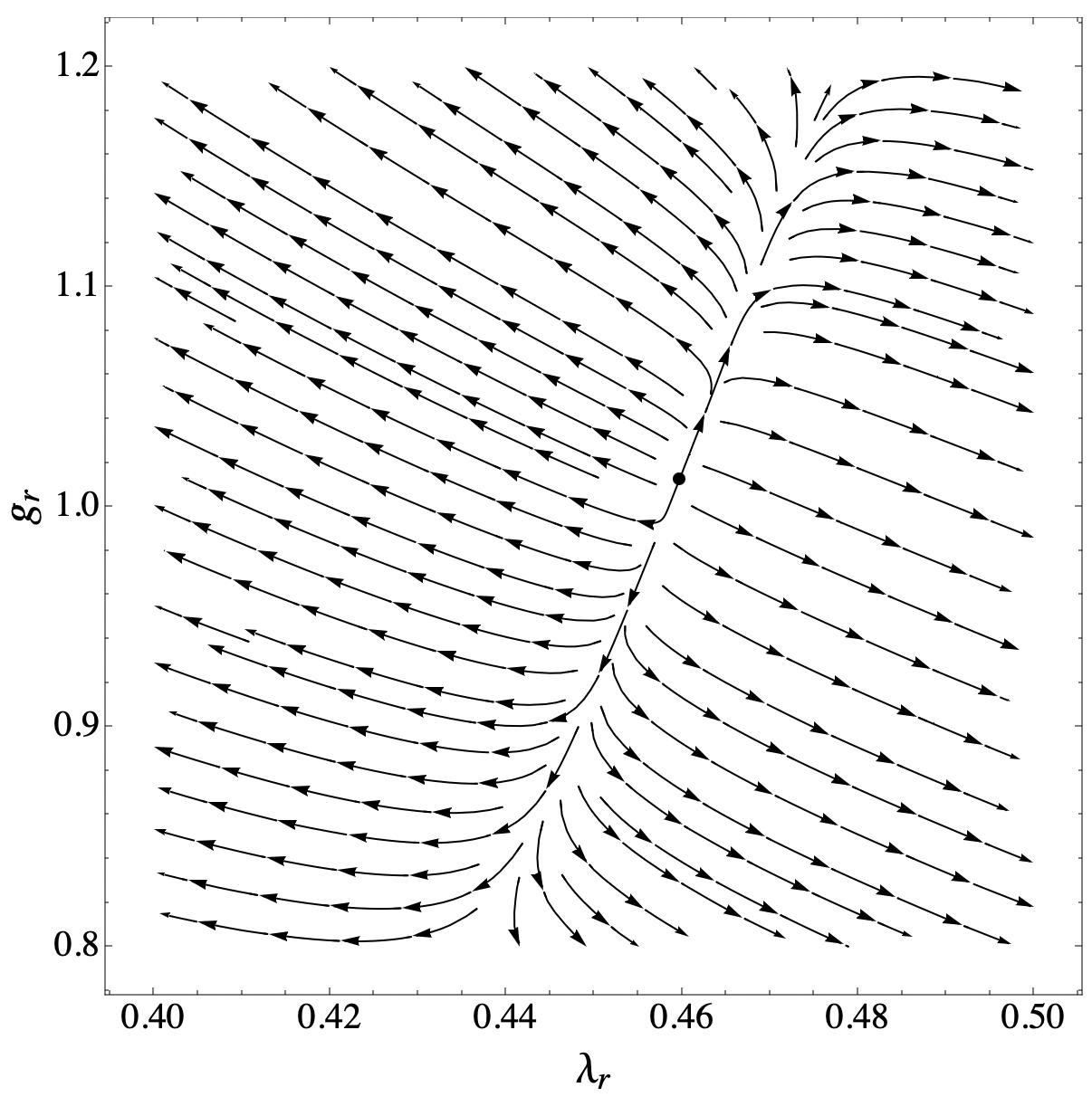}\quad \includegraphics[width=.49\textwidth]{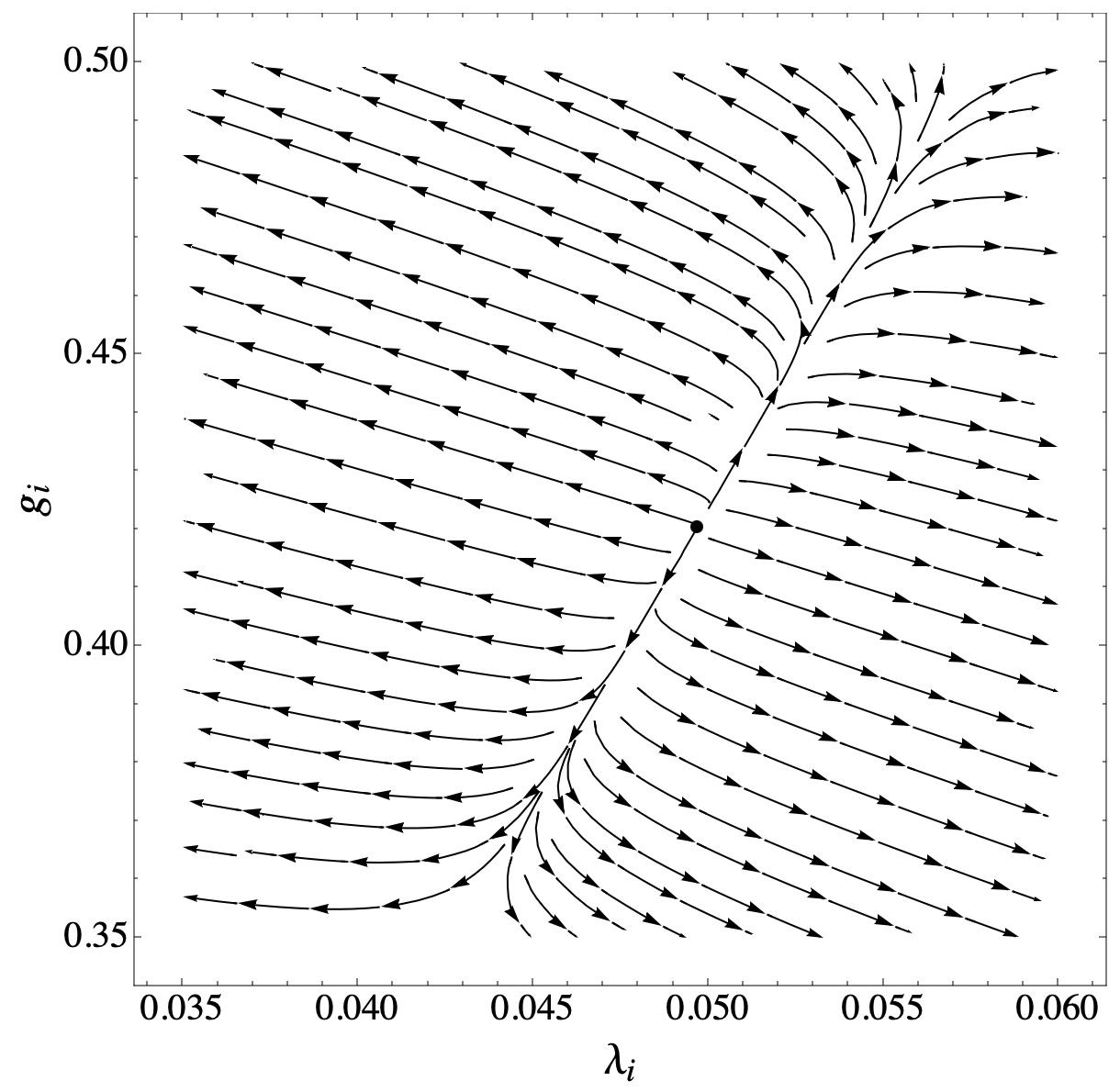}
	\caption{Case (a), four scalar fields \cite{Ferrero:2024rvi}. Projection of the real ($\lambda_r,\; g_r$) and the imaginary part ($\lambda_i,\; g_i$) of the flow near the UV fixed point (black dot) at $ (\lambda_*,g_*)= (0.45962 + 0.0496779\; i,1.0127 + 0.420496\; i)$. The arrows point towards decreasing $k$.}
	\label{fig:2}
\end{figure}

With all the ingredients at hand, let us briefly recap the main findings: the coupling constant related to the matter contribution $\kappa(k)$ is found not to flow and also  not to affect the flow of the gravitational coupling in the truncation considered here. Furthermore, due to the imaginary heat kernel coefficients as in \eqref{HKLorentz}, the flow of the coupling constants is complex: the two gravitational coupling constants $G_N(k)$ and $\Lambda (k)$ have complex UV fixed point (see Figure \ref{fig:2}).

The  properties of the trajectories are investigated and it is found that there exist trajectories which have real dimensionful coupling constants in the IR, when $k \to 0$. These were called \textit{admissible trajectories}  \cite{Ferrero:2024rvi}.

\bigskip

In the case of the Gaussian dust matter (b) - which can be  interpreted as a system of free test particles scattered throughout space - an interesting feature arises: due to the particularly simple structure of the physical Hamiltonian, after having integrated out the conjugate momenta 
one can employ Euclidean methods  by performing a controlled analytic continuation.  Importantly, this still describes a Lorentzian signature quantum gravity model.
The Hamiltonian formulation connects to the Euclidean path integral via a
Wick rotation of the time-evolution operator $e^{i H T} \to e^{-HT}$. This is well defined, since
the dust frame provides a preferred time $T$, avoiding ambiguities in defining an imaginary time \cite{Baldazzi:2018mtl,Banerjee:2024tap}.
 In fact, the synchronous gauge ensures no dynamical lapse/shift, making the analytic continuation $t \to -it$ mathematically consistent.
Hence, the reduction performed on the Gaussian dust  simplifies the canonical structure, in such a way to enable a direct mapping to Euclidean methods to derive the generating functional of the Schwinger $n$-point functions. Effectively, the Gaussian dust matter’s reference frame circumvents the usual obstructions to Wick rotation in GR. By Higgsing the diffeomorphism group, the theory retains Lorentzian causality at the quantum level while exploiting Euclidean techniques for computational tractability.

In this second investigation, the authors accounted for the non-trivial measure by implementing an appropriate field redefinition (see Ref. \citen{Falls:2025sxu} for a recent discussion on the meaning of field redefinitions)
\begin{equation}
\mathcal{D}q\; \mathcal{M}[q] \quad \to \quad \mathcal{D}Q\;.
\end{equation}
This procedure allowed  to absorb the complexities of the measure into the new field variables $Q$, thereby preserving the correct physical content of the theory.

In this case, the path integral is formulated over an ADM-foliated spacetime and cannot be recast as an integration over the full spacetime metric $g$. The partition function is
\begin{equation}\label{eq:38}
	Z_\text{GD}[j] =\int\; [\mathcal{D}Q]\; 
 \overline{\Omega[Q(\infty)]}\; \Omega[Q(-\infty)]\;e^{-S[Q]} \; e^{\int\; \text{d}^dx \; j^{ab} q_{ab}}\;,
\end{equation}
with the action
\begin{equation}\label{eq:39}
S[g(Q)]=-\frac{1}{16 \pi G_N} \int \; \text{d}^dX\; \sqrt{\det(g(Q))}\left(R[g(Q)]-2\Lambda\right) \;.
\end{equation}
Here $g(Q)$ is the Euclidean signature 
spacetime metric constructed from $q(Q)$ with $X$ restricted to be  synchronous coordinates, i.e.,
obeying the Euclidean line element 
\begin{equation}
g_{\mu\nu}(X)\; \text{d}X^\mu\;\text{d}X^\nu=(\text{d}x^0)^2+[q(Q))]_{ab}(X)\;\text{d}x^a\; \text{d}x^b\;.
\end{equation}
Such a restriction is a direct consequence of the deparametrization performed by means of the dust fields.
The action under such a restriction can also be rewritten as
\begin{equation}
		\begin{aligned}
S[q(Q)]=&-\frac{1}{16 \pi G_N}\int \text{d}t \int\;\text{d}^{d-1} x \; \sqrt{q(Q)}\left((q(Q)^{ac} \; q(Q)^{bd}-q(Q)^{ab} \; q(Q)^{cd})\;\frac{\dot q_{ab}\;\dot q_{cd}}{4}\right.\\
	&\left.\qquad\qquad\qquad\qquad\qquad\qquad-\left(R^{(d-1)}[q(Q)]-2\Lambda\right)\right)\;.
	\end{aligned}
\end{equation}

It is also important to mention that we neglected the Gibbons-Hawking boundary contribution in \eqref{eq:39}, which naturally appeared in the construction of the path integral \cite{Ferrero:2025idz}.

The ansatz for the EAA in this case results in
\begin{equation}\label{eq:55}
	\begin{aligned}
	\Gamma_k[g]&=-\frac{1}{16 \pi G_N(k)} \int \text{d}t\int\;\text{d}^{d-1} x \; \sqrt{q(Q)}\left((q(Q)^{ac} \; q(Q)^{bd}-q(Q)^{ab} \; q(Q)^{cd})\;\frac{\dot q(Q)_{ab}\;\dot q(Q)_{cd}}{4}\right.\\
	&\left.\qquad\qquad\qquad\qquad\qquad\qquad\qquad-\left(R^{(d-1)}[q(Q)]-2\Lambda(k)\right)\right)\;.
	\end{aligned}
\end{equation}
We notice that no further gauge-fixing freedom is left here, since the dust  fields have exhausted all the gauge redundancies from the gravitational part.

In this setting, Euclidean methods become applicable, bringing the computation closer to standard techniques used in AS analyses. In particular, within the context of AS significant progress has been made in recent years toward developing FRG methods that address renormalization on foliated spacetimes \cite{Rechenberger:2012dt,Saueressig:2023tfy,Korver:2024sam, Saueressig:2025ypi}.
The major challenge in this context lies in the loss of access to standard covariant heat kernel techniques, which are central to conventional computations in FRG. Instead, one must perform more intricate and technically demanding evaluations, often requiring refined methods that explicitly account for the anisotropies between spatial and temporal components which might have been introduced by the foliation \cite{Contillo:2013fua,DOdorico:2015pil,Barvinsky:2021ubv,Barvinsky:2024kgt}.

\begin{figure}[h]
	\centering
	\includegraphics[width=.77\textwidth]{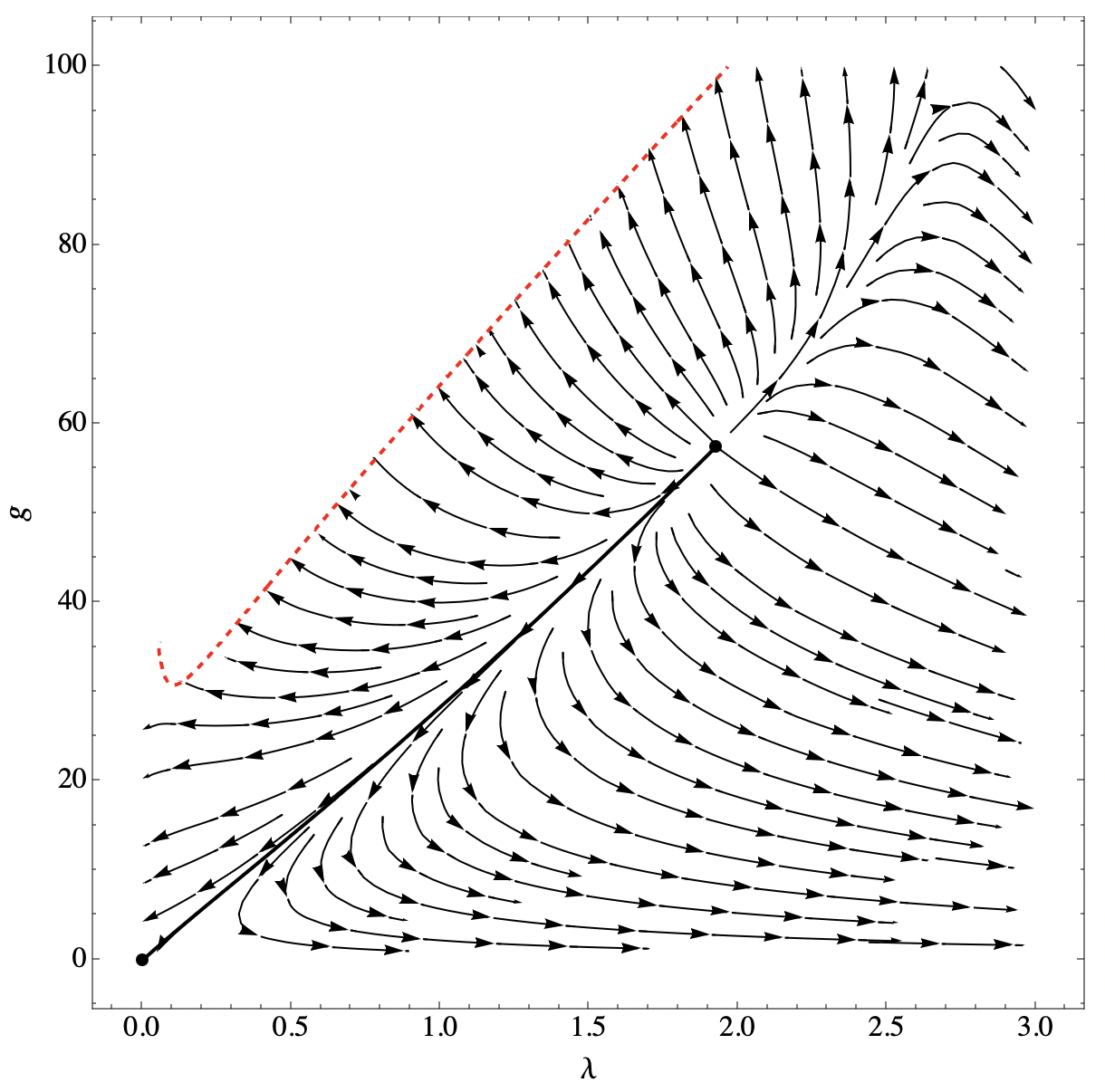}
	\caption{Case (b), Gaussian dust \cite{Ferrero:2025idz}. The trajectories stem from the UV fixed point and flow towards decreasing $k$. The thick line is the trajectory ending at $k = 0$. The dashed red line represents the zone after which the flow becomes singular.}
	\label{fig:3}
\end{figure}

The model in eq. \eqref{eq:55} exhibits a UV fixed point $(\lambda_*, g_*)=(1.92 ,  57.41)$ and critical exponents ($\theta_1 = 8.01, \theta_2 =2.13$). These are in qualitative agreement with the AS computations which  are also based  on the Lorentzian ADM decomposition \cite{Saueressig:2025ypi}, even though the methods and the chosen cutoff are different.  In Figure \ref{fig:3} we report the flow diagram and we depict the UV and IR fixed points.
Importantly a new versatile tool was used, imported from loops computations in Feynman integrals: the Mellin-Barnes integrals \cite{10.1112/plms/s2-6.1.141,Czakon:2005rk,Jantzen:2012cb,Smirnov:2012gma}. This allowed a controlled evaluation of the higher order contributions in the proper time integration (see Appendix B in Ref. \citen{Ferrero:2025idz}). 

\subsection{Novel features from asymptotically safe canonical quantum gravity}
In the preceding subsection, we presented two concrete realizations of \textit{asymptotically safe canonical quantum gravity}. We now turn to a comparative analysis, linking these results to the broader setup developed in Section \ref{sec:4}. Our aim is to identify and examine the novel features that emerge in the renormalization group flows associated within these two examples, highlighting their distinctive characteristics and the implications they carry for the asymptotic safety scenario in the canonical framework.

From a technical point of view, no gauge-fixing term is added to the ansatz of the effective average action (hence no ghosts are involved). Nevertheless, in the case of the scalar fields we emphasize how a rewriting of the residual gauge freedom in terms of a gauge-fixing might be  possible. Furthermore, different features of the heat kernel techniques are used. In the case of the scalar fields we exploit for the first time the Lorentzian heat kernel, while in the Gaussian dust case these techniques are inspired by the ADM-foliated flows \cite{Rechenberger:2012dt,Saueressig:2023tfy,Korver:2024sam, Saueressig:2025ypi}.

Importantly, the measure in this context is determined by the underlying canonical model and is not a free choice. This is natural, as different choices of measure would correspond to different deparametrizations of the theory, namely different choices of relational time \cite{Bonanno:2025xdg}.

 Furthermore, also the initial and the final states affect the properties of the RG of the theory. This is particularly important in  Lorentzian theories, where no unique vacuum state is given on a curved spacetime \cite{DAngelo:2023wje}.

\bigskip

In the first case, which is worked out in Lorentzian signature, the renormalization group flow exhibits complex behavior. This was something completely new and unusual within the AS framework and comes with a number of new open questions. While this may initially appear unconventional, it is not without precedent in a broader QFT setup: imaginary contributions to the effective action have been previously considered in the context of Schwinger pair production in quantum electrodynamics \cite{Schwinger:1951nm,Dittrich:1985yb}, where they signal physical phenomena such as vacuum instability and particle creation. Similarly, also in a gravitational setting such complex contributions have been discussed \cite{Fecit:2025kqb,Boasso:2024ryt,Akhmedov:2024qvi,Akhmedov:2024axn,Maldacena:2024spf}, particularly in cosmological scenarios \cite{Marolf:2010zp}. The connection towards this physical picture is an intriguing topic left for future investigations.
This point also highlights the inherent ambiguity in defining a RG flow in Lorentzian signature. In such a setting, the regularization of the d'Alembertian operator becomes subtle: for certain field configurations, its eigenvalues can become negative, which in turn leads to an effective complexification of the corresponding momentum modes. This complicates the standard interpretation of scale separation and mode integration that underlies the RG flow.
The choice of cutoff can, of course, influence this property. In this initial analysis, the authors focused on developing a mathematically stable framework for reaching the well-defined limit $k \to 0$, without attempting to interpret the integration over Lorentzian modes. Exploring alternative, more physically motivated cutoff choices remains an important direction for future work.

Furthermore, a deeper understanding of the interplay between the chosen truncation and the resulting complex structure is needed. In our setting, the complex nature of the coefficients originates from the heat kernel expansion, which is truncated at a finite order to match the structure of the EAA. Similar imaginary contributions  also appeared in perturbative one-loop computations \cite{Callan:1977pt,Weinberg:1987vp}.  This raises the question of whether the observed complex contributions are merely artifacts of the approximation or truncation scheme.

Moreover, the flow does not present the typical spiraling and complex conjugated critical exponents as those presented in Section \ref{sec:4}. Finally, due to numerical approximations needed to solve the flow detailed in Ref. \citen{Ferrero:2024rvi}, the trajectories do not reach the Gaussian fixed point. However, this result arises solely as an artifact of numerical approximation and does not reflect a fundamental property of the system.
On the other side, there is no singular region for positive $\lambda >1/2$, meaning that those trajectories going towards $\lambda\to \infty$ for decreasing $k$ can be integrated down to $k \to 0$, reaching the effective regime.

\bigskip

As the Gaussian dust is concerned, the used RG techniques are Euclidean and on a foliated spacetime. The flow presents a real fixed point, in qualitative good agreement with the ADM foliated asympototic safe flows. The critical exponents are both positive and real, signaling the presence of two relevant directions. Finally, due to the choice of a regulator in proper time (see eq. \eqref{cutoffpt} in Section \ref{sec:4}), the region which cannot be trusted anymore is on the upper left side of the first quadrant (cf. Figure \ref{fig:3}), meaning that the trajectories with $\lambda>0$ along the entire flow in $k$ can be integrated down to the IR. This is an important feature to make connection to observations, because it allows access to those trajectories which are believed to be compatible with current cosmological observations, namely with an accelerated expansion of the universe ($\Lambda>0)$ \cite{SupernovaSearchTeam:1998fmf,Planck:2018vyg}. We emphasize that in many AS analyses the singular region does not allow this investigation and those trajectories are not complete.

To facilitate the integration in proper time, a powerful new method has been introduced: the use of Mellin–Barnes integrals. This technique enables a systematic and efficient evaluation of higher-order integrals involving various combinations of cutoff functions and propagators. Its versatility makes it particularly well-suited for computing proper-time integrals with different regulator choices, providing a direct and flexible approach to evaluating non-perturbative flow equations in the proper-time formalism.

\bigskip

It is also important to emphasize that the models consider have a different number of degrees of freedom than the two of pure gravity. For instance, if one considers the Gaussian dust, there are six propagating physical degrees of freedom corresponding to two gravitational and four scalar propagating modes. Therefore, the proposed material deparametrization alters the physical content, raising the question of how it can be meaningfully compared to pure gravity. On the other hand,  since pure gravity may not be directly observable in practice, this approach could still offer valuable physical insights.

Finally, this investigation opens new intriguing questions regarding the nature of the RG. It appears that the relational nature of the underlying theory translates into a \textit{relational RG}. As we mentioned in the section before, a given universality is associated with critical phenomena.
{Indeed, for all we know, quantities derived via a relational strategy in a particular physical frame will generally depend on that frame, and it is not yet fully understood what aspects of the dynamics, such as the RG flow, remain invariant under a change of physical frame, or what class of transformations relates observables defined in different frames. These transformations are not diffeomorphisms by construction, since the reference fields themselves carry physical meaning. Clarifying their nature and the possible equivalence of frames is therefore an important open issue, though not one that implies any tension with  covariance or diffeomorphism invariance. In approximate regimes involving idealized reference systems, neglecting backreaction and quantum properties, the physical frame effectively corresponds to a specific gauge or coordinate choice, and in such cases the expected gauge invariance and frame equivalence of results are recovered.}

{In particular, in the reduced framework discussed here, where the gauge degrees of freedom are discarded from the beginning, one can question whether different reference frames would correspond to the same universality class.} Although this is something expected once the proper transformation between reference frames has been taken into account, it might be challenging to realize, especially in a reduced phase space. Moreover, in certain situations where the reference frames are ``non-ideal" \cite{delaHamette:2021oex} and/or the causal structure of the underlying quantum spacetime exhibits specific features, some frames may be better suited than others for detecting the observables (see Refs. \citen{Frob:2017gyj,Frob:2021ore,Frob:2023awn} for some examples in a perturbative cosmological setting). On a speculative level, this could imply that, in such ``non-ideal" material reference frames, the theory may fail to exhibit a fixed point. These aspects open up promising directions for future research and will be explored in more detail in forthcoming work.

\section{Conclusions and Outlook}	\label{sec:7}
In this paper, we have argued how Asymptotic Safety and Canonical Quantum Gravity are not opposing approaches to quantum gravity, but rather can be combined to highlight different aspects of the ultimate theory of quantum gravity. Both approaches respect the principle of background independence and employ non-perturbative methods. Their synergy, particularly in the effort to define viable candidates for observables, can provide valuable insights.

The interface between the gravitational Lagrangian and Hamiltonian plays a crucial role. We analyzed their differences and the assumptions needed to transition from one formalism to the other. In general,  for the full theory of quantum gravity, the notion of gauge-fixing is different for the two setups, making a complete equivalence not transparent. Despite this, there exist models in which the connection can be made more explicit, allowing the use of methods typically associated with the opposite formalism. This can be achieved for instance in a relational framework.

We have highlighted how certain aspects of the path integral can be adapted  in order to account for a canonical theory of quantum gravity, by relaxing explicit covariance and defining the path integral in a relational manner. The path integral and in particular the effective action, was then constructed by means of the FRG setup. The flow equation is used as a  tool to reach the $k \to 0$ limit, where all the quantum fluctuations have been integrated out.

It is expected that this relational, non-covariant path integral can be connected to different relational, non-covariant path integrals through suitable transformations that account for changes of the observer. This is a mathematically challenging task and should be explored in future work. Such non-covariance is not necessarily troubling since it arises from the physical characteristics of observables and observers in GR and as long as transformations between reference systems remain well-defined. This can potentially be achieved at the level of the unreduced phase space by leveraging the framework developed in the context of quantum reference frames \cite{Loveridge:2017pcv,Vanrietvelde:2018pgb,AliAhmad:2021adn} {and TGFT cosmology \cite{Gielen:2016dss,Oriti:2016qtz}}.
On the other hand, we have outlined in Section \ref{sec:3} how  initial steps have been made toward defining a covariant theory from Hamiltonian quantization. This approach could serve as a foundation for future work, aiming to bridge the two frameworks and apply standard AS techniques to the resulting path integral.

Along this analysis, new technical and formal aspects of the path integral construction have been highlighted, including the measure, the state dependence, and the signature. In order to make the connection between the two approaches more direct, it is compelling to use Lorentzian flow equations. This paves the way to new methods and interpretational aspects, such as the appropriate choice of the cutoff function, the (non-)uniquess of the flow, and the state dependence.

These novel features of asymptotically safe canonical quantum gravity present a new challenge, but also a new opportunity to advance both quantum gravity approaches and testing the ultimate theory of quantum gravity against observations. 
As a promising avenue, for instance, these new methods open the possibility to compute by means of non-perturbative RG tools correlations function of cosmological perturbations in a relational manner \cite{Giesel:2007wi,Giesel:2007wk,Frob:2017lnt}. Apart from material reference frames, an important tool to exploit are so-called geometrical clocks, which have a interpretation in terms of standard cosmological gauges \cite{Giesel:2018opa}.

\section*{Acknowledgments}
I am  grateful to Thomas Thiemann for fruitful discussions and to Alfio Bonanno, Muxin Han, Roberto Percacci  and especially  Frank
Saueressig for fruitful discussions and   comments on
the manuscript. I thank Daniele Oriti for valuable comments that clarified the link between canonical and covariant quantum gravity models and the relational strategy’s compatibility with general covariance.
I am supported by the Friedrich-Alexander University (FAU) Emerging Talents Initiative (ETI).

\section*{ORCID}
\noindent Renata Ferrero - \url{https://orcid.org/0000-0003-0613-1274}



\bibliographystyle{ws-ijmpa}
\bibliography{sample}
\end{document}